\documentclass{aa}

\usepackage{graphicx,natbib}
\usepackage[varg]{txfonts}
\usepackage{amsmath,amssymb,color}
\bibpunct{(}{)}{;}{a}{}{,}

\begin{document}

\title{Non-stoichiometric amorphous magnesium-iron silicates in circumstellar dust shells}

\subtitle{Dust growth in outflows from supergiants}

\author{
  Hans-Peter Gail\inst{1}
 \and  Akemi Tamanai\inst{2}   
 \and Annemarie Pucci\inst{3}
 \and Ralf Dohmen\inst{4}
}

\institute{
Institut f\"ur Theoretische Astrophysik,  Zentrum f\"ur Astronomie, Universit\"at Heidelberg, 
           Albert-Ueberle-Str. 2,
           69120 Heidelberg, Germany, email:~{\tt gail@uni-heidelberg.de}
\and
RIKEN Cluster for Pioneering Research, Star and Planet Formation Laboratory, 
2-1 Hirosawa, Wako-shi, Saitama 351-0198, Japan
\and
Universit\"at Heidelberg, Kirchhoff-Institut f\"ur Physik,
           Im Neuenheimer Feld 227,
           69120 Heidelberg, Germany
\and
Ruhr-Universit\"at Bochum, Institut f\"ur Geologie, Mineralogie und Geophysik,
           Universit\"atsstr. 150,
           44780 Bochum, Germany
  }


\date{Received date ; accepted date}

\abstract{}
{
We study the growth of dust in oxygen-rich stellar outflows in order to find out to which extent dust growth models can quantitatively reconcile with the quantities and nature of dust as derived from observations of the infrared emission from circumstellar dust shells.  
}
{We use a set of nine well-observed massive supergiants with optically thin dust shells as testbeds because of the relatively simple properties of the outflows from massive supergiants, contrary to the case of AGB stars. Models of the infrared emission from their circumstellar dust shells are compared to their observed infrared spectra to derive the essential parameters that rule dust formation in the extended envelope of these stars. The results are compared with a model for silicate dust condensation.
}
{For all objects, the infrared emission in the studied wavelength range, between $6\ \mu$m and $25\ \mu$m, can be reproduced rather well by a mixture of non-stoichiometric iron-bearing silicates, alumina, and metallic iron dust particles. For three objects ($\mu$ Cep, RW Cyg, and RS Per), the observed spectra can be sufficiently well reproduced by a stationary and (essentially) spherically symmetric outflow in the instantaneous condensation approximation. For these objects, the temperature at the onset of massive silicate dust growth is of the order of 920 K and the corresponding outflow velocity of the order of the sound velocity. This condensation temperature is only somewhat below the vapourisation temperature of the silicate dust and suggests that the silicate dust grows on the corundum dust grains that formed well inside of the silicate dust shell at a much higher temperature. The low expansion velocity at the inner edge of the silicate dust shell further suggests that, for these supergiants, the region inside the silicate dust shell has an only subsonic average expansion velocity, though a high degree of supersonic turbulence is indicated by the widths of spectral lines. 
}
{Our results suggest that for the two major problems of dust formation in stellar outflows, that is (i) formation of seed nuclei and (ii) their growth to macroscopic dust grains, we are gradually coming close to a quantitative understanding of the second item. }



\maketitle

\section{Introduction}

At the end of their lives, stars return much of their initial mass to the interstellar medium (ISM), either by long-lasting massive stellar winds or by short eruptive or explosive events. These mass-return processes are almost inevitably accompanied by the condensation of the abundant refractory elements into small solid particulates, and this freshly produced dust is injected into the ISM. The outflowing matter evolves hereby from a transparent gas near the star into an opaque gas-dust mixture at some distance. The dust is heated by the radiation field of the central star and becomes the source of an intense thermal radiation in the near- to mid-infrared spectral region. This infrared emission from warm dust in the vicinity of highly evolved stellar objects is intensively studied by observations and theoretical modelling because one can derive in this way a great deal of information on the mass-loss processes of stars and of their role as putative sources of interstellar dust. 

The details of the mechanisms by which matter is returned by stars are different for low- and intermediate-mass stars (initial mass $M\lesssim 8$ $M_{\sun}$) and massive stars (initial mass $M\gtrsim 8$ $M_{\sun}$). Low- and intermediate-mass stars finally arrive at the end of their lives on the asymptotic giant branch (AGB), where they become pulsationally unstable, while the higher mass stars between $ 8\ {\rm M}_{\sun}\lesssim M \lesssim 40\ {\rm M}_{\sun}$ end their lives on the red giant branch (RGB), where they are also prone to a slight variability, but not to strong pulsations. In both cases, the stars develop intense stellar winds with ever increasing  mass-loss rates as they climb up on the AGB or RGB and increasingly develop optically thick dust shells. Massive stars with initial masses $M\gtrsim 40\ {\rm M}_{\sun}$ do not arrive at the RGB -- they remain too hot for dust formation in a stellar wind -- but they probably all eject large amounts of mass in short outburst events with accompanying intense dust formation as they become luminous blue variables (LBV) \citep{Smi14,Smi17}.
 
The mechanism of dust formation depends critically on the density and temperature of the gas out of which the solids condense. The thermodynamic state of the ejected matter is therefore one important factor that has to be accounted for if one starts to study dust formation in a stellar outflow. The thermodynamic state is mainly determined by the properties of the flow, and this differs significantly between the AGB and the RGB. For stars on the AGB, the complicated non-linear coupling between dust formation and the hydrodynamics of pulsation \citep[see][]{Gai13} makes it difficult to clearly separate between the different roles that dust condensation and hydrodynamics play in the final observable effects. While for carbon-rich element mixtures one effectively has no choice but to study dust formation in common with pulsation because cool carbon stars are only found on the AGB, for dust formation in oxygen-rich element mixtures the massive stars seem to present a more favourable case. Their net outflow properties seem to be simpler since one does not have to deal with the strong shocks running through the atmosphere and the higher zones of the star as is the case for AGB stars. 

\begin{table*}

\caption{Sample of supergiants and their basic data.}

\begin{tabular}{lccclcrclrrrl}
\hline
\hline
\noalign{\smallskip}
     & Spectral & Variable & Period\tablefootmark{a} & class\tablefootmark{g} & $\Delta V\tablefootmark{h} $ & $T_\mathrm{eff}$\tablefootmark{b} & \multicolumn{1}{c}{$L$\tablefootmark{b}} & \multicolumn{1}{c}{$\dot M_\mathrm{gas}$\tablefootmark{c}} & $v_\mathrm{exp}$\tablefootmark{c} & $M$ & $D$\tablefootmark{d} & $A_V$\tablefootmark{b} \\
\noalign{\smallskip}
Name & type\tablefootmark{a}  & type\tablefootmark{a} & \multicolumn{1}{c}{[d]} &  &  & \multicolumn{1}{c}{[K]} & \multicolumn{1}{c}{[L$_{\sun}$]} & \multicolumn{1}{c}{[$\rm M_{\sun}a^{-1}$]} & [km/s] & $M_{\sun}$ & [pc] & \\
\noalign{\smallskip}
\hline
\noalign{\smallskip}
$\mu$ Cep & M2Ia     & SRc & 730  & 2.SEc & 1.2  & 3\,700 & $3.4\times10^5$  & $5.0\times10^{-6}$ & 20 & 15 & 870 & 2.01 \\
RW Cyg    & M3Iab    & SRc & 550  & 2.SEc & 1.5  & 3\,600 & $1.45\times10^5$ & $3.2\times10^{-6}$ & 23 & {\it 15} & 1\,320 & 4.49 \\
W Per     & M3Iab    & SRc & 485  & 2.SEc & 2.6  & 3\,550 & $5.5\times10^4$  & $2.1\times10^{-6}$ & 16 & {\it 15} & 1\,900 & 4.03 \\
RS Per    & M4Iab    & SRc & 245  & 2.SEc & 1.8  & 3\,550 &  $1.4\times10^5$ & $2.0\times10^{-6}$ & 20 & {\it 15} & 2\,400 & 2.63 \\[.3cm]
$\alpha$ Ori & M2Iab & SRc & 2335 & 2.SEc & 0.9  & 3\,800 & $5.6\times10^4$ &  $1.5\times10^{-6}$ & 15 & {20}  & 130    & 0.62 \\
$\alpha$ Sco & M1.5Iab & Lc &     & 2.SEc & 1.0  & 3\,550 & $7.1\times10^4$ &  $1.0\times10^{-6}$ & 17 & 18 & 185    & 0.72 \\
S Per     & M4.5Iab  & SRc & 822  & 3.SE  & 4.5  & 3\,500 & $8.6\times10^4$ &  $7.5\times10^{-6}$ & 20 & 20 & 1\,900 & 2.63 \\
PZ Cas    & M3Iab    & SRc & 925  & 2.SEc & 2.0  & 3\,605 & $1.9\times10^5$ &  $8.3\times10^{-6}$ & 30 & {\it 15}  & 2\,510 & 2.11 \\
VX Sgr    & M5/M6III & SRc & 730  & 3.SE  & 5.8  & 3\,700 & $3.4\times10^5$ &  $5.0\times10^{-6}$ & 20 & 10 & 1\,570 &      \\
\noalign{\smallskip}
\hline
\end{tabular}
\tablefoot{
\tablefoottext{a}{Taken from the SIMBAD database.} \tablefoottext{b}{\citet{Lev06}.} \tablefoottext{c}{\citet{Mau11}.} \tablefoottext{d}{\citet{Jon12}.} \tablefoottext{e}{\citet{Com16}.}  \tablefoottext{f}{\citet{VBe09}.}  \tablefoottext{g}{\citet{Slo03}.} \tablefoottext{h}{\citet{Kis06}.}

}

\label{TabStarParm}
\end{table*}

For this reason we studied the type and quantities of dust formed in massive supergiants that are presented in \citet[henceforth called paper I]{Tam17}. We presented in that paper newly determined optical constants for non-stoichiometric iron-bearing silicates of intermediate composition between olivine-like and pyroxene-like materials and applied them to model the dust content of dust shells by fitting model spectra to observed infrared spectra for a set of four supergiants. Such investigations of supergiant spectra have been previously performed \citep{Ver09,Spe00} based on other sets of optical data for the possible dust materials. From studies of pre-solar dust grains by \citet{Vol09a}, \citet{Bos10}, \citet{Ngu10}, and \citet{Bos12}, it appears more likely, however, that the silicates formed in circumstellar environments have a composition more similar to the materials we have studied in the laboratory, that is, they have an intermediate composition between amorphous olivine-like and pyroxene-like silicates. Here we extend our previous work to a larger set of supergiant stars for a comparison and to provide more details on the dust formation process in such objects. 

As previous studies have shown, three different major dust components are formed in supergiants (and probably also in the AGB stars): silicate minerals, as witnessed by the two broad solid state emission bands around $\approx 10$ $\mu$m and 18 $\mu$m, aluminium oxide dust, witnessed by a broad emission feature around $\approx 12$ $\mu$m, and iron dust, witnessed by an infrared continuum emission seen between $\approx 6$ and $\approx 9$ $\mu$m where no other likely material existing in an oxygen-rich environment could contribute. 

From these main dust species, aluminium oxide dust is thought ever since the eariest studies to condense at higher temperature than the silicate dust \citep[for instance ][]{Sal77}. Models for circumstellar dust shells confirm this in so far, as the fit of model spectra to observed spectra requires the inner edge of the aluminium oxide dust to be closer to the star and to show a  higher dust temperature than the inner edge of the silicate dust shell. This means that we have to deal with two different possibilities for condensation of silicate dust: 

First, the silicate dust precipitates on top of the already existing aluminium oxide dust once temperature has sufficiently dropped such that the silicates become stable against evaporation. During subsequent further cooling the condensible material is rapidly consumed from the gas phase and silicate dust exclusively condenses around preformed aluminium oxide particles. The onset of silicate dust condensation then is expected to occur close to the stability limit of the silicate.

This does not mean that each silicate particle has a detectable aluminium oxide core, which could in principle be identified in pre-solar dust grains. Until now, the investigated pre-solar silicate dust grains \citep{Vol09a,Bos10,Ngu10,Bos12} seem not to show indications for a frequent occurrence of such a phenomenon, but since aluminium oxide and silicates form to a limited extent a solid solution, the initial core may dissolve in the silicate mantle as it is overgrown and looses its identity.

Second, the silicate dust initially starts to precipitate on top of the aluminium oxide dust, but the subsequent growth is slow such that not much of the condensible material from the gas phase is consumed until the temperature dropped to the limit where direct avalanche nucleation of silicate dust from the gas phase commences and growth on these seed particles takes over as the dominating silicate condensation process. The onset of massive silicate dust condensation then occurs at a temperature significantly below the stability limit of silicate material because significant degrees of supersaturation are required for onset of nucleation. This case has been studied, for instance in our paper \citet{Gai13a}.   
 
The aim of the present study is to derive information on the condensation temperature of silicate dust and on the growth efficiency of the silicate material. These are important bricks of information for further studies of the silicate dust formation process in an oxygen-rich environment. 

We study also an additional problem: the formation of metallic iron dust particles as a separate dust component. The Fe atoms from the gas phase may either be incorporated into the silicate material as cations within the lattice structure of olivine- or pyroxene-like minerals or of a material of intermediate composition as studied in paper I. Or the Fe atoms form a separate solid phase of metallic iron. From analysis of infrared emission from dust shells one knows that the Fe is found in both phases, in silicates and as metal. With respect to the metallic component one has to consider two prospects. Either the iron  forms a completely separate dust component, or it mainly forms as inclusions within the silicate dust grains (or both). In interplanetary dust particles an abundant component are Grains with Embedded Metal and Sulphides (GEMS) where one finds that nano-sized particles of metallic iron and iron sulphide are embedded in a glassy silicate matrix. Some fraction of them are circumstellar condensates according to their unusual isotopic abundances of oxygen \citep{Kel11}. This shows that at least part of the iron is found as tiny inclusions in silicate grains. We aim in this paper also to study how much of this kind of `dirty' dust is formed and at which temperature. 

\section{Comparison stars}

Infrared emission from circumstellar dust in the environment of galactic supergiants has been investigated for a moderate large sample of such objects by \citet{Spe00} and \citet{Ver09} in order to study common features and individual differences of the dust properties in the sample. In our paper I we selected from these studies and from the atlas of \citet{Slo03b} four objects with optically thin circumstellar dust shells which satisfy the requirement that there is a sufficiently strong contribution of dust to the mid-infrared emission that can be used to analyse the nature and quantities of dust formed by such stars. Optically thick sources were excluded, however, to avoid difficulties to disentangle dust properties and radiative transfer effects. Objects with obvious contributions from crystalline dust were also omitted. Here we look for objects that satisfy the same requirements and the additional requirement that for all objects the outflow velocity and the mass-loss rate of their stellar wind has been determined from analysis of molecular or atomic lines. This additional information enables us to relate dust mass-loss rates determined by modelling of the infrared emission to gas mass-loss rates and to determine condensation efficiencies. \citet{Mau11} have listed the supergiants where mass-loss rates have been determined by spectroscopic means. From this compilation we add five further objects to our previous samples, such that our final sample contains \object{$\mu$ Cep}, \object{RW Cyg}, \object{W Per}, \object{RS Per}, \object{$\alpha$ Ori}, \object{$\alpha$ Sco}, \object{S Per}, \object{PZ Cas}, and \object{VX Sgr}. Table \ref{TabStarParm} lists all objects used in the present study and some key data taken from literature that are required for our modelling.

In terms of the classification scheme by \citet{Kra02} the selected objects all correspond to type 2.SEc or 3.SEc (in a few cases). ISO-spectra for these objects are taken from the atlas of fully processed spectra from the SWS\footnote{%
{Downloaded from \tt https://users.physics.unc.edu/{\textasciitilde}gcsloan/ 
library/swsatlas/atlas.html}}
described in \citet{Slo03}. These spectra cover a wavelength range from 2.4 $\mu$m to 45.4 $\mu$m. Since in many cases the long wavelength part $\lambda>25$ $\mu$m of the spectral data from this catalogue is very noisy we only use the spectral region $\lambda<25$ $\mu$m for our analysis. Since none of the selected objects show indications for the presence of significant amounts of crystalline dust material, all spectral features of interest from dust are from amorphous dust which are all located within the spectral region $9\,\mu\mathrm{m}\lesssim\lambda\lesssim25\,\mu\mathrm{m}$. Since we use for our comparison between observed and model spectra only the mid-infrared spectrum we do not correct the observed spectra for interstellar absorption. We checked that as described in paper I, but found in all cases the required corrections to be insignificant for the wavelength range of interest.

\subsection{Stellar properties}

The basic stellar parameters required for constructing of a model of its dust shell, $T_\mathrm{eff}$ and $L$, are taken from \citet{Lev06}. For our sample of stars the surface element abundances seem not to have been determined for the individual stars, except for $\alpha$ Ori, and then also only for the elements C, N, and O. Therefore, we can only use typical element abundances for massive supergiants for our investigations on dust formation. Massive supergiants on the RGB represent the final stage in the life of a massive star. Because of their short lifetime, they formed initially with an element mixture found in the present-day interstellar medium. Therefore, they show, without any known exception, an oxygen-rich element mixture. This initial mixture underwent some modifications by rotational driven mixing and the dredge-up process at the base of the RGB where material from the nuclear burning central region is mixed into the surface layers. Significant modifications are observed, however, only for the light elements C, N, O. The main dust species formed in their outflow are basically composed of only the few highly abundant and refractory rock-forming elements Mg, Al, Si, Ca, Fe, and Ni in compounds with O. Other chemical elements may condense into minor dust species or may be included in the main species as impurities. 

For oxygen its abundance excess over the rock-forming elements and carbon is sufficiently high that even after part of the oxygen is consumed by CO formation there remains much more than what could at most be bound in solids ($\epsilon_\mathrm{O}-\epsilon_\mathrm{C}\gg2\epsilon_\mathrm{Si}+\epsilon_\mathrm{Mg}+1.5\epsilon_\mathrm{Al}+\epsilon_\mathrm{Ca}+\epsilon_\mathrm{Fe}+\epsilon_\mathrm{Ni}$). Under these circumstances the precise value of the oxygen abundance is not important for the dust formation because there is always more than enough oxygen available to oxidise all rock-forming elements if required. The abundance of the rock-forming elements remains unchanged from stellar birth up to the RGB stage because no material is mixed to the surface that suffered the required high temperatures for affecting their abundances by nuclear burning processes. 

\begin{table}

\caption{Photospheric element abundances of the dust forming elements in the
astrophysical logarithmic scale for different groups of stars. The data are from 
\cite{Lod09} for F and G stars, and from \citet{Nie12} for B stars.}

\begin{tabular}{l@{\hspace{.9cm}}lll@{\hspace{.5cm}}r}
\hline\hline
\noalign{\smallskip}
    &\multicolumn{2}{c}{ F \& G stars}&& \multicolumn{1}{c}{B dwarfs}\\
El. & \multicolumn{1}{c}{solar $Z$} & age $<1$\,Ga & \\
\noalign{\smallskip}
\hline
\noalign{\smallskip}
He &               &               && $10.99\pm0.01$ \\
C  & $8.37\pm0.11$ & $8.39\pm0.11$ && $8.33\pm0.04$ \\
O  & $8.75\pm0.07$ & $8.77\pm0.13$ && $8.76\pm0.05$ \\
Mg & $7.63\pm0.32$ & $7.64\pm0.21$ && $7.56\pm0.05$ \\
Al & $6.52\pm0.24$ & $6.54\pm0.22$ &&  \\
Si & $7.60\pm0.28$ & $7.61\pm0.23$ && $7.50\pm0.05$ \\
Ca & $6.42\pm0.37$ & $6.48\pm0.39$ &&  \\
Fe & $7.55\pm0.12$ & $7.61\pm0.25$ && $7.52\pm0.03$ \\
Ni & $6.22\pm0.09$ & $6.25\pm0.07$ &&               \\
\noalign{\smallskip}
\hline
\end{tabular}

\label{TabElAbu}
\end{table}

For our considerations on dust formation we, therefore, use element abundances determined for B stars from the solar vicinity of the Milky Way as reported by  \citet{Nie12} which are probably the most accurate ones presently known. Because of their youth the B stars represent the composition of the present ISM from which also the presently observed massive supergiants formed. For the elements not analysed in that paper (Al, Ca, Ni) we use data from \citet[ Table 11]{Lod09} for young F and G stars (age $<1$ Ga). Values are given in Table~\ref{TabElAbu}. 

We note that for B stars as well as for young F and G stars the abundance of Fe is comparable to or slightly higher than the Si abundance. This is different from what one observes for the solar system element mixture and in particular for abundances in low- to intermediate-mass giant stars of lower metallicities where silicon has significantly higher abundance than Fe. This means that for dust formation in the massive giants Fe plays a more important role than for the less massive and much older AGB stars.   

For  modelling the stellar outflow we need to know stellar masses. What we need is the mass on the RGB, not the initial mass. This can only be determined directly if the red giant is a member of a binary system. For our comparison stars this is only met by $\alpha$ Sco for which \citet{Rei08} give a mass of 18 $M_{\sun}$ for the cool supergiant in the system. In other cases the mass can be estimated by comparison with stellar evolutionary tracks. For $\mu$ Cep an initial mass of 15 to 20 $M_{\sun}$ is determined by \citet{Mon19} in this way. We assume a mass of 15 $M_{\sun}$ for the RGB because some mass is lost during the previous evolution. For $\alpha$ Ori \citet{Ohn14} and \citet{Dol16} give an initial mass of $20\pm5$ $M_{\sun}$ and $20^{+5}_{-3}$ $M_{\sun}$, respectively. We take a value of 20 $M_{\sun}$ as current mass. For S Per a mass of 20 $M_{\sun}$ was estimated by \citet{Yat94} by an analysis of observed expansion velocities of maser spots, but this is probably very uncertain. For VX Sgr its variability properties resembles more a Mira variable than a massive supergiant \citep{Chi10a}, but are not really typical for a Mira, and might represent a massive supergiant after extensive mass-loss. We provisionally assume a current mass of 10 $M_{\sun}$. For all other stars of our sample we assume a mass of 15 $M_{\sun}$ which may be a typical value for the current mass of massive supergiants on the RGB \citep{Bea20}. The masses used in our calculations are given in Table~\ref{TabStarParm}. All these stellar masses are rather uncertain, but fortunately they only weakly enter into calculations of the stellar outflow (cf. Sect.~\ref{SectWiMo}). 


\section{Dust opacity}

\label{SectAbsCo}

\subsection{The dust components}

\label{SectDustComp}

We consider in our model calculation the three most abundant dust materials that have been detected in dust shells of massive supergiants: amorphous silicates, amorphous aluminium oxide, and metallic iron. Their emission dominates the mid- to far-infrared emission from dust shells. Additional dust species of minor abundance are certainly present, but we neglect them because they are not crucial for the main characteristics of the infrared spectrum.

Aluminium oxide is thermal much more stable than silicate material and therefore is generally assumed to exist in a stellar outflow already at higher temperature than silicates. In this case there exists the possibility that the silicate material grows on top of pre-existing aluminium oxide particles.  
Here we assume that the aluminium oxide dust and the silicate dust exist as separate dust particles. The possibility that they form composite grains with an oxide core and a silicate mantle is not considered. Such composite grains have been found in small number as pre-solar dust grains \citep[see for instance ][]{Lei18} while in contrast to this pre-solar silicate grains without recognizable foreign cores are abundant. The rarity of such composite grains does not require, however,  that precipitation of silicates on aluminium oxide cores is only a minor route to silicate dust formation because silicate and aluminium oxide may form a solid solution. 

The presence of metallic iron dust in circumstellar shells of supergiants was inferred from an observed continuous excess emission in the wavelength range $5\ \mu\mathrm{m}\lesssim\lambda\lesssim8\ \mu\mathrm{m}$ where the mineral dust components are highly transparent \citep{Ver09,Tam17}. An alternative possibility would be the emission of a warm chromosphere which has also been proposed as a source of this continuum emission \citep{Gil74}. For our stellar sample, $\alpha$ Ori and $\alpha$ Sco shows clear evidence for having a chromosphere \citep{Lin17,OGo20}. For all others there seems to be no evidence of a chromosphere.

Presently it is unclear whether this iron exists as a separate dust component or if small iron particles are included into a growing silicate matrix. Such nanometre-sized inclusions are postulated to be responsible for the observed absorptivity of `astronomical silicates' \citep{Oss92} and have been detected in pre-solar GEMS (= Glass with Embedded Metal and Sulphide) from oxygen-rich stars \citep{Kel11}. The possibility that pre-solar iron grains may exist as separate particles seems never to have been investigated. 

In any case, we have to consider both possibilities in our model calculations: that there exists an iron-bearing silicate dust component without inclusions of iron nano particles and an iron-bearing silicate with iron inclusions. This allows us to handle the situation that solid iron is thermally less stable than silicate material such that first a silicate material is formed without iron inclusions and that only at lower temperatures there may (or may not) appear a material with such inclusions. Hence, for treating the silicate-iron-complex in the opacity calculation we include three dust materials in our model: iron-bearing silicates, silicates with iron inclusions, and pure iron grains. 

\subsection{Particle shape and size}

\label{SectSizeShape}

For the present study we have selected stars that form amorphous silicate dust. The formation of amorphous silicate material in a cooling gas has been studied in the laboratory for silicate material with forsterite-like composition and for amorphous SiO$_2$ \citep{Kam05,Kam05b,Kim08,Ish15}. The particles were produced by vaporisation of mixed powders of Mg and SiO in an O$_2$-Ar-atmosphere and subsequent seed particle formation and particle growth from the vapour. Final particle sizes were up to about 200 nm, that is, just the sizes relevant also for circumstellar dust. The particles collected from the smoke all have a clear spherical or almost spherical shape, even if they are finally crystallized. Additionally, industrial production of nano-particles in flames shows that amorphous nano-particles have always spherical shape \citep[see, for example, ][ for a review]{Kam01}. Though formation conditions of dust particles in the laboratory and in circumstellar environments are different  with respect to density, gas-phase composition, and cooling timescales there is no recognizable reason why the shape of particles formed by seed particle formation and growth in a stellar outflow should be different if the material has amorphous structure.

Contrary to this, experience with radiative transfer calculations for circumstellar dust has shown that with spherical grains and laboratory-determined optical constants of dust analogues one cannot fit the profiles of the 10 $\mu$m and 18 $\mu$m bands of silicates. The width of the profiles is too small and centred on a too short wavelength. The most likely origin of this discrepancy is a non-spherical shape of the grains. A good fit to observed profiles of the silicate bands is usually obtained with the popular continuous distribution of ellipsoids (CDE) approximation \citep[see][]{Boh83}. Studies on the extinction properties of small non-spherical particles \citep[for instance ][]{Min03,Min05,Min07} and laboratory experiments \citep{Tam09} have shown that the deviation of the profiles calculated with CDE from profiles obtained with other kinds of asymmetric particle shape distributions is much smaller than their respective deviations from the profile calculated for spheres. A usage of the CDE approximation, thus, accounts for the essential effects resulting from a non-sphericity of the particles without requiring that this has to be taken literally as an approximation for the real particle shape distribution. In this sense we will use the CDE approximation for calculating dust absorption as a typical representative of the class of non-spherical grains. 

The necessity to use CDE for obtaining reasonable fits to observed profiles means that the dust particles mainly contributing to the infrared emission in the wavelength region of the silicate bands are definitely not spherically symmetric. This is in conflict with the laboratory results, that amorphous particles that nucleate from the gas phase are born as spheres. The most likely explanation for this discrepancy seems to be that the particles, once they grow to sufficient size as to decouple dynamically from the gas, start to drift through the gas and agglomerate to clusters via particle-particle collisions. This process seems never to have been studied for dust in circumstellar shells. According to the crude estimate given in \citet[section 6.8.5]{Gai13} the number of collision that a particle suffers with another particle during its residence time at distances $<10^{17}$ cm is at most a few, that is, only cluster with a small number of members can be formed which nucleated from the gas phase and grew initially as spherical particles. Since such aggregates are subject to continued vapour deposition growth, complex shaped particles are likely to emerge at some distance from the inner edge of the dust shell. The experimental (and theoretical) study of \citet{Tam18} of the extinction properties of non-spherical particles formed by aggregates of a small number (2 to 4) of SiO$_2$ spheres demonstrate how the optical properties of aggregates differ from that of their spherical constituents. The trends for the differences are similar to the differences between the result for spheres and CDE calculations.

The applicability of the CDE approximation assumes that the particle size (here and in the following meaning radius of a sphere with equal volume) is small compared to the considered wavelength. In analyses of dust emission from giant stars it was found that the radius of dust grains is typically of the order of 0.1 $\mu$m. This is suggested by observations of polarisation of light from dusty stars \citep[for instance ][]{Sha75,Mau86} and by spatial extension of the cloud of OH masers around supergiants \citep{Jur96}. In analysing the mid-infrared emission from circumstellar dust shells one considers wavelengths $\lambda\gtrsim 5\ \mu$m and in particular the wavelength range $9\ \mu\mathrm{m}\lesssim\lambda\lesssim20\,\mu\mathrm{m}$ containing the two silicate features.  Even if the dust grains of some object should be unusually large there is ample room to apply the CDE approximation to that wavelength range. 

\subsection{Optical constants}

\subsubsection{Aluminium oxide}

Optical constants of the amorphous aluminium oxide for the mid- to far-infrared wavelength region $\lambda\ge7.8\ \mu$m are taken from \citet{Beg97} (the constants for porous amorphous aluminium oxide). For shorter wavelengths they are augmented by data from \citet{Koi95}. 

\subsubsection{Silicate}
\label{SectOptConSil}

For the silicate dust we use in the mid- to far-infrared wavelength region the optical constants for the amorphous non-stoichiometric magnesium-iron silicates from paper I. We obtained with them very good fits between model spectra and observed spectra for the four objects considered in paper I, both for the 10 $\mu$m and the 18 $\mu$m feature. It appears that the optical constants of this kind of material are representative for the optical constants of the silicate material in oxygen-rich supergiants. 

Our optical data are for an amorphous material derived by pulsed laser deposition from four target materials with composition Mg$_2$SiO$_4$, Mg$_{1.6}$Fe$_{0.4}$SiO$_4$, Mg$_{0.8}$Fe$_{1.6}$SiO$_4$, and Fe$_2$SiO$_4$. The composition of the resulting condensates (denoted as Si-film1, \dots, Si-film4 in paper I) is shown in Table 1 of paper I. The Mg\# of the deposit, defined as the atomic number ratio Mg/(Mg+Fe) of the material, is different from that of the target materials, but also spans the range from 0 to 1. For a stellar object we do not know the Mg\# in advance and have to determine it as part of the fitting procedure between the stellar and model spectrum. We interpolate for this purpose for an assumed value of Mg\# the optical constants linear between the corresponding adjacent tabular values and look for the best match between model and observation. Hence the optical constants for the iron-bearing silicates are determined from interpolation in a table. How we constructed the table is described in paper I. For convenience, we add here the resulting data set for $2\ \mu\mathrm{m}\lesssim\lambda\lesssim200\mu\mathrm{m}$ as Table \ref{TabOptDat4} in Appendix~\ref{AppOptTab}.

For calculating dust temperatures we need data for the optical constants in the wavelength range $0.4\ \mu\mathrm{m}\lesssim\lambda\lesssim 4\ \mu\mathrm{m}$. Unfortunately we were not able to do the corresponding measurements on our probes in the optical and near infrared wavelength range. For wavelength shorter than $2\ \mu$m we have to augment our optical constants by data from other sources. In the model calculations it turns out that the spectra are best fitted by silicates with Mg\#\ between that of Si-film2 and Si-film3; data for the material Si-film1 and Si-film4 are actually not used. Therefore, we have to augment our tabular data for the two cases Si-film2 and Si-fim3.

In case of Si-film2 we augment our data in the optical to near-infrared spectral region by forsterite data from \citet{Sco96}. These data refer to a material which was produced as an amorphous film deposited on a cold substrate by a similar PLD method as used by us. Under these circumstances, the resulting material should  also be the same kind of non-stoichiometric material as obtained in our experiments, though no details of the composition and structure of the material have been given by \citet{Sco96}. Additionally, \citet{Wad03} obtained an amorphous, non-stoichiometric silicate by vapour deposition on a cold substrate.  Since \citet{Sco96} used as target material a (unspecified) natural forsterite -- which always have a low, but non-vanishing iron content -- their amorphous material in any case is not pure forsterite but also has a small iron content. Unfortunately, no value for this was reported. We use the \citet{Sco96} data in the wavelength range $\lambda<4\ \mu$m. The cut between both data sets is chosen at $4\ \mu$m because at this wavelength the real part of the complex index of refraction is almost identical for both data sets and the slight discrepancy of the imaginary part is inconsequential because the silicates are highly transparent in this region.

In case of Si-film3 no optical data for the optical and near-infred spectral region of an iron-rich amorphous silicate fabricated by vapour deposition on a cold substrate could be found. The sole published optical data in the relevant wavelength region for amorphous iron-bearing silicates seems to be the data of \citet{Dor95}. Their material was synthesised by quenching of a melt of stoichiometric composition. The structure and composition of this material most likely deviates from the kind of material with composition intermediate between olivine and pyroxene as obtained by vapour deposition. Because of a lack of better data, we augment our data for Si-film3 for $\lambda<4\ \mu$m by the data from \citet{Dor95} for their olivine-like material with~$x=0.5$. In this case there is a significant jump of the imaginary part $k$ at $\lambda=4\ \mu$m in the resulting data set. Again, this is inconsequential since the silicate material is highly transparent in this region and practically no contribution to the resulting dust spectrum in this region comes from the silicates.  

If instead of the optical constants used in this work the frequently used data set for amorphous, iron-bearing silicate with Mg\# = 0.5 from \citet{Dor95} is used for the whole wavelength region, it is also possible to obtain reasonable model fits for the emission from the dust shells of our set of comparison stars, in particular for the two strong silicate bands. The main difference is that the deviation between the model spectrum and the observed one is up to 10\% in the absorption trough between the two strong silicate bands for the \cite{Dor95} data while it stays below $\sim3$\% for our data.

\subsubsection{Iron}

For the metallic iron we use tabulated optical data from \citet{Ord88}. A pair of Drude models is fitted to these data because the conduction band of transition metals like iron is the overlap of two bands stemming from the atomic 4s and 3d electron states such that two bands are involved in the conduction process \citep{Mot35}. This model is used to extend the tabulated data for wavelengths outside the range of the table of Ordal. 

With respect to absorption by metallic iron we consider two possibilities, either that there exists a separate iron dust component, or that iron is included as nano-sized particulates in silicate dust. The second case is approximated by using  an effective medium theory for the optical constants of a heterogeneously composed material \citep[cf.][]{Boh83}. Mixing theories are often used to calculate optical properties of metal-insulator composite materials \citep[see for instance ][ for a short discussion on the applicability and limits of mixing rules to such materials]{San10}. We calculate effective optical constants for the composite material via the Bruggeman mixing rule because this usually provides acceptable results for all mixing ratios of the components \citep[see also][]{Vos07}. 

In this approximation one calculates $\epsilon_\mathrm{eff}$ from the equation  \citep[cf. the review of ][]{Ber95}
\begin{equation}
\epsilon_{\rm eff} = \left( \sum^{N}_{i=1} \frac{3f_i}{2\epsilon_{\rm eff} + \epsilon_i} \right)^{-1}\,,
\label{EffMediumBrugg}
\end{equation}
where $\epsilon_i$ are the complex dielectric functions of the components, $f_i$ their volume fractions in the composite material, and $N$ is the number of components. For a small volume fraction $f\lesssim0.3$ of iron inclusions in a silicate matrix this results in a kind of dirty dust material with strongly enhanced absorption in the near- and mid-infrared spectral region were silicate dust is usually transparent. Such an enhancement of absorptivity of cosmic dust was postulated in the first quantitative modelling of the spectrum of oxygen-rich circumstellar dust shells by \citet{Jon76}, where the `dirty dust' model for cosmic silicate dust was introduced for the first time, and is discussed in detail in \citet{Oss92}. For higher fractions of iron the optical properties of the mixture are completely dominated by the iron component.

We chose as matrix material an iron-poor silicate because the opacity outside the two strong silicate resonances is completely dominated by the iron inclusions and the iron content of the matrix plays a subordinate role for the opacity of the mixture. 

\begin{table}

\caption{Constants used in the calculation of the dust opacity and assumed maximum grain radius}

\begin{tabular}{lllllll}
\hline
\hline
   & $\rho_\mathrm{sol}$ & $A_\mathrm{sol}$ & $\epsilon_\mathrm{c}$ & $\nu_\mathrm{key}$ & $a_\mathrm{max}$ \\
Species & [g\,cm$^{-3}$] &   &  &  [$\mu$m] \\
\noalign{\smallskip}
\hline
\noalign{\smallskip}
Fo40    & 3.45 & 178.8 & Si & 1 & 0.2 \\
Fo80    & 3.45 & 153.6 & Si & 1 & 0.2 \\
Iron    & 7.87 & 55.85 & Fe & 1 & 0.1 \\
Alumina & 3.97 & 51.0  & Al & 2 & 0.2 \\
\noalign{\smallskip}
\hline
\end{tabular}

\label{TabCoSu}
\end{table}

\subsection{Absorption and scattering coefficients}

The calculation of the absorption and scattering coefficients from the optical constants follows standard assumptions with respect to shape and size of the circumstellar dust grains. First, with respect to the shape of the particles, we assume that the dust particles either are small spheres with radius $a$, or they are small ellipsoids with a distribution of axis ratios corresponding to the CDE approximation and an equivalent radius $a$. Second, we assume that the dust grains are from a distribution of radii, $f(a)$, between some lower and upper radius $a_\mathrm{min}=5$ nm and $a_\mathrm{max}$, respectively. For the size distribution we make the conventional assumption that $f(a)$ equals the size distribution given by \citet{Mat77} (denoted as MRN distribution), that is, $f(a)=Na^{-7/2}$ with a normalisation constant, $N$, such that the integral of $f(a)$ extended between $a_\mathrm{min}$ and $a_\mathrm{max}$ equals one. The value of $a_\mathrm{max}$ used in most of our calculations is given in Table~\ref{TabCoSu}.

\begin{figure}

\includegraphics[width=\hsize]{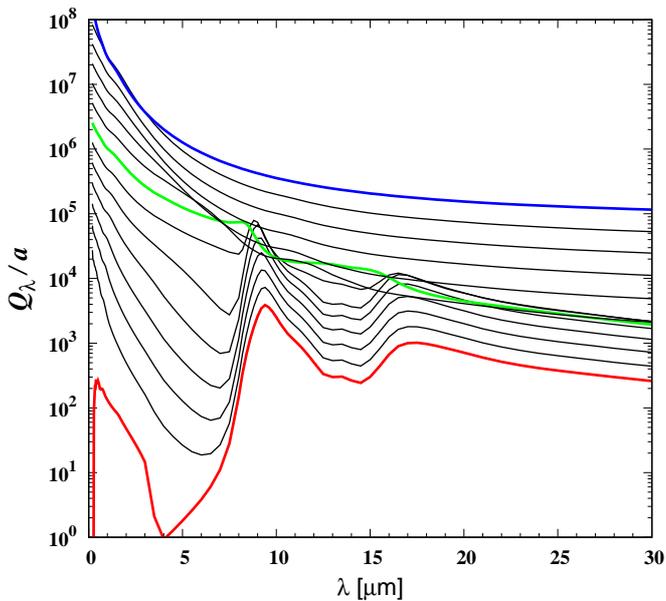}

\caption{Absorption efficiency $Q_\lambda$ for a composite material of forsterite and metallic iron with varying volume fractions of iron from 0\% to 60\% in steps of 5\%. The red line corresponds to pure forsterite, the green line to a volume fraction of iron of 35\% at the border between insulator-like and metal-like absorption behaviour, and the blue line corresponds to pure iron. For clarity, successive lines are vertically shifted by a factor of two with respect to each other.  }

\label{FigIronIncl}
\end{figure}

The absorption, extinction, and pressure cross sections normalised to the geometric cross section $\pi a^2$ of a particle, the efficiencies $Q_\lambda^\mathrm{a,\,e,\,pr}$,  then are calculated either by means of Mie theory for spherical grains \citep{Boh83} or by the model denoted as continuous distribution of ellipsoids \citep[CDE, see][]{Boh83}. The mass-absorption, $\kappa_\lambda^{\textrm{a}}$, and mass-extinction, $\kappa_\lambda^{\textrm{e}}$,  coefficients with respect to unit mass of dust material, and the corresponding quantity for the radiation pressure, $\kappa_\lambda^{\textrm{pr}}$,  are calculated from
\begin{equation}
\kappa_\lambda^{\textrm{a,\,e,\,pr}}={3\over4}{1\over\rho_{\mathrm{sol}}}
\int_{a_\mathrm{min}}^{a_\mathrm{max}}{Q_\lambda^\mathrm{a,\,e,\,pr}\over a}\,a^3f(a)\,\mathrm{d}a\ \mbox{\Large/} 
\int_{a_\mathrm{min}}^{a_\mathrm{max}}a^3f(a)\,\mathrm{d}a
\,,
\label{DefDustKap}
\end{equation}
where $\rho_{\mathrm{sol}}$ is the mass-density of the dust. To obtain the corresponding mass absorption and extinction coefficients with respect to unit mass of the dusty gas this has to be multiplied by the dust-to-gas ratio
\begin{equation}
\psi={A_\mathrm{sol}\epsilon_\mathrm{c}\over1+4\epsilon_\mathrm{He}}\,,
\label{DefDuGasFac}
\end{equation}
where $A_\mathrm{sol}$ is the atomic weight corresponding to the chemical formula of the dust material, $\epsilon_\mathrm{c}$ the element abundance of the element which is considered to be key for the condensation of the considered dust material (usually this is Si for silicate minerals, Al for aluminium oxide dust, and Fe for iron dust), and $\epsilon_\mathrm{He}$ the element abundance of He in the dusty gas (assumed to be dominated by H and He). Numerical values used in our calculations are shown in Table~\ref{TabCoSu}.

This all refers to the case where the key element is completely condensed into the dust species. If only some fraction is condensed, $\kappa_\lambda^{\textrm{a,\,e,\,pr}}$ has to be multiplied by the fraction $\,f_\mathrm{c}$ of the key element actually condensed into dust.

\begin{figure}

\includegraphics[width=\hsize]{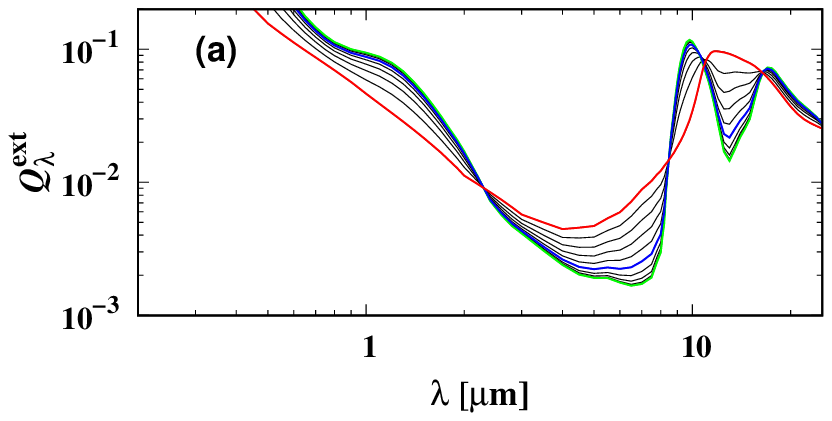}
\includegraphics[width=\hsize]{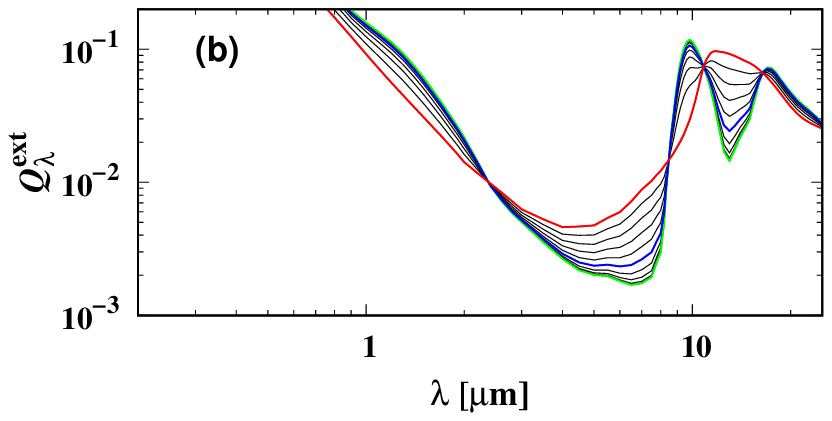}

\caption{
(a) Extinction efficiency $Q_\lambda^\mathrm{ext}$ for composite spherical particles of silicate grown on the surface of alumina grains, calculated for core-mantle grains with fixed total radius $a=0.1\ \mu$m and different core-radii from 0 to 1 of the total radius in steps of 0.1 (black lines). The green and red lines correspond to pure silicate and alumina grains, respectively. The blue line is for a grain with a core of 0.5 of the total radius. (b) Superposition of $Q_\lambda^{\mathrm{ext}}$-values for pure silicate and pure alumina grains weighted according to the volume portions in the core-mantle grains.
}

\label{FigAluPliCM}
\end{figure}

\subsection{Silicate dust with iron inclusions}

In Fig.~\ref{FigIronIncl} the dependence of the absorption efficiency $Q_\lambda$ for a heterogeneous iron-forsterite mixture on the volume fraction of iron in the mixture is shown. For the silicate component we used the optical constants of the sample Si-film2 as described in Sect.~\ref{SectOptConSil}.

For constructing the figure we calculated from the optical constants of the silicate and of iron via the Bruggeman mixing rule for given volume fractions of iron the effective optical constants of the composite material, via Mie theory the absorption efficiency $Q_\lambda$, and then averaged with respect to a MRN size distribution. Volume fractions of iron between 0\% and 60\% in steps of 5\% are chosen for the computation. For low iron content ($\lesssim30$\%) the absorption by the iron inclusions mainly fills-up the absorption gaps between the optical region and $\lambda\approx 8\ \mu$m and between the two silicate peaks, but does not much change the general appearance of the silicate absorption. For iron volume fractions $\gtrsim 40$\% the absorption equals the absorption by pure iron; the forsterite inclusions within the iron matrix are no more visible from the outside. In between (around $\approx35\%$), there occurs a significant change in the spectral variation of the absorption of the composite material, corresponding to a transition from an insulator-like behaviour to a metallic-like behaviour.

\subsection{Core-mantle grains}

Since solid aluminium oxide is much more refractory than silicate minerals, it is to be expected that in a stellar outflow at least part of the silicate material is deposited on the surface of alumina grains. Figure \ref{FigAluPliCM}a shows for the crucial wavelength range from 5 to 25 $\mu$m important for comparison of models for dust shells with observations the extinction efficiency for such core-mantle grains, calculated by Mie-theory \citep[using the code BHCOAT from ][]{Boh83}. The total grain radius is set to 0.1 $\mu$m and the core radius is varied from zero to the full radius. A radius of 0.1 $\mu$m could be typical for dust grains formed in massive supergiants \citep{Jur96}. Figure \ref{FigAluPliCM}b shows the sum of the extinction efficiencies of pure silicate and alumina grains weighted with the volume frations as in the core-mantle grains. The two results are very similar. In particular for the mid-infrared spectral region where models are usually compared with observations the opacities calculated for both cases are essentially the same. We take this as a justification to treat in the opacity calculation silicate and alumina dust as separate components, though in reality the silicate dust may, at least in part, form as a coating on the alumina dust.


\section{Model of the dust shell}

\label{SectWind}

\subsection{Wind model}

\label{SectWiMo}

We extend the pure radiative transfer model for the circumstellar dust shell used in paper I by an approximate treatment of the outflow properties of the stellar wind. The real structure of the outflow is most likely very complex due to the giant convection cells existing in the atmospheres of massive giant stars \citep[cf.][]{Fre02,Chi09,Chi10}. In the stellar atmosphere and the overlaying layers the flow is probably dominated by these convection cells with in- and outflows varying on scales of somewhat less than the stellar radius while farther out it turns to the systematic outflow observed at large distances. The true mechanism initiating the outflow from the star is still unknown. In any case, however, once dust formed by the most abundant refractory elements (that is silicate dust in our case) starts to condense in a cooling outflow around a luminous star, the radiation pressure on dust and the close momentum coupling between dust and gas rapidly accelerates the dusty gas to  highly supersonic outflow velocities. Most of the infrared emission from the warm dust heated by stellar radiation originates in this acceleration zone of the flow because the dust density is highest there. The near infrared emission spectrum from an optically thin dust shell therefore contains mainly information on this zone of wind acceleration and initial dust growth. For analysing the infrared emission from the dust it suffices to consider in zero order accuracy to care only about the flow properties in this zone and farther outwards and leave aside the complicated conditions in the inner zone where the flow pattern is complex because of giant convection cells and possibly also by shocks. 

In this acceleration zone the outflow velocity is of order of the sonic velocity or supersonic such that pressure effects are of subordinate importance. We then restrict ourselves to the very simple approach that has initially been proposed by \citet{Geh71} and has been used since then in many papers and is discussed in detail, for instance, in \citet{Fer06}: We assume a spherically symmetric, stationary, and supersonic outflow where pressure effects are neglected. The radial variation of the outflow velocity, $v$, then is determined by the equation
\begin{equation}
v\,{\textrm{d}\,v\over\textrm{d}\,r}=-{GM_r\over r^2}+{4\pi\over c}\,\varkappa_{\mathrm{pr}}H\,,
\label{EqVel}
\end{equation}
where only the gravitational attraction by the star and the radiative acceleration of the gas-dust mixture is considered. $M_r$ is the mass contained within a sphere with radius $r$. For simplicity, a tight dynamical coupling between gas and dust  is assumed. The tight coupling approximation is not valid at very low mass-loss rates \citep[$\lesssim10^{-7}\ {\rm M}_{\sun} a^{-1}$, see][]{Gai13}, but none of our sample stars has a really low mass-loss rate, though $\alpha$ Ori and $\alpha$ Sco are at the verge of applicability. 

\begin{figure}

\includegraphics[width=\hsize]{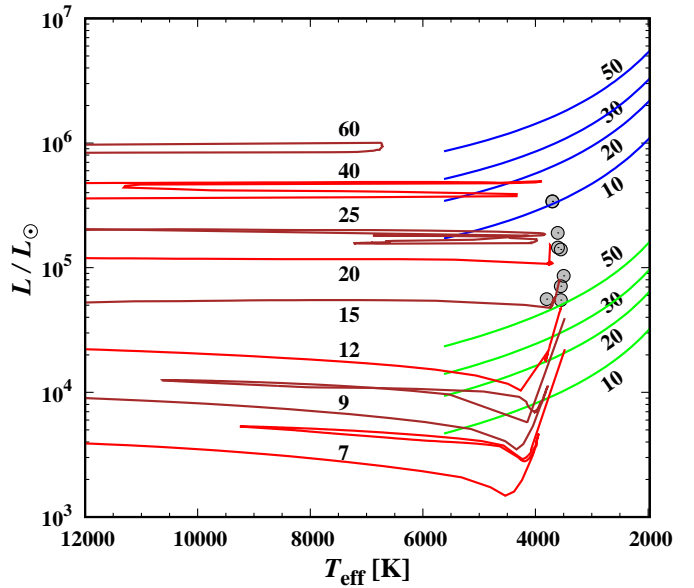}

\caption{Hertzsprung-Russell diagram with evolutionary paths for non-rotating massive stars that enter at the end of their live the RGB \citep[from][]{Eks12}. The grey dots correspond to the set of supergiants from Table~\ref{TabStarParm}. The lower set of lines (green) in the right part of the figure shows the Eddington luminosity, Eq.~(\ref{DefEddLum}), calculated for silicate dust and for stellar masses of 10, 20, 30, and 50 $M_{\sun}$. The corresponding upper set of lines (blue) is for alumina dust. These limits are derived with the assumption that all Si or Al is condensed into their respective dust components. For a lower fraction $f$ of condensation the lines have to be scaled upwards~$\propto1/f$.   }

\label{FigHR}
\end{figure}

We consider in this paper optically thin dust shells that contain only small amounts of mass compared to the stellar mass and approximate $M_r$ by the stellar mass $M_*$. In the radiative acceleration term, $\varkappa_\mathrm{pr}$ is the flux average of the extinction coefficient for radiation pressure per unit mass of the gas-dust mixture and $4\pi H$ the total energy flux of the star for which we have
\begin{align}
H&=H_*\,R_*^2/r^2\\
4\pi\,H_*&=\sigma_{\mathrm{SB}}T_\mathrm{eff}^4\\
L_*&=4\pi r^2\,4\pi H\,.
\end{align}
Here $L_*$ is the stellar luminosity, $T_\mathrm{eff}$ its effective temperature, $\sigma_\mathrm{SB}$ the Stefan-Boltzmann constant, and $R_*$ the stellar radius.

The quantity $\varkappa_{\mathrm{pr}}$ is given by
\begin{equation}
\varkappa_{\mathrm{pr}}={1\over\varrho}\sum_i \kappa_{1,i}^{\mathrm{pr}}\,\varrho_i=\sum_i\kappa_{1,i}^{\mathrm{pr}}\psi_i\ f_i\,, 
\label{DefKapPrMix}
\end{equation}
where the index $i$ refers to the dust species. The quantity $\kappa_{1,i}^{\mathrm{pr}}$ is calculated according to Eq.~(\ref{DefDustKap}) from $Q_\lambda^{\mathrm{pr}}$ under the assumption that the key element for dust species $i$ is fully condensed into that species and is averaged with respect to the stellar energy flux spectrum. The quantity $f_i$ is the fraction of the key element of dust species $i$ actually condensed into that dust species. The conversion factor $\psi_i$  from mass absorption per unit dust mass to unit mass of the gas-dust mixture is defined by Eq.~(\ref{DefDuGasFac}). We always calculate $Q_\lambda^{\mathrm{pr}}$ by means of Mie theory because the maximum flux of the stellar radiation peaks in the wavelength range 1\,$\mu$m \dots\ $1.5\ \mu$m where the wavelengths are not sufficiently large compared to the dust particle radius that the small-particle approximation can safely be applied.
 
Integrating Eq. (\ref{EqVel}) from the inner edge $r_0$ of the silicate dust shell to some radius $r$ results to
\begin{equation}
v^2-v_0^2=-v_\mathrm{esc}^2\left(1-{r_0\over r}\right)+{L_*\over 2\pi c}\sum_i\kappa_{1,i}^{\mathrm{pr}}\psi_i\ \int_{r_0}^r {f_i\,\mathrm{d}r\over r^2}\,, 
\label{EqWindVel}
\end{equation}
where $v_0$ is the initial velocity with which the gas flows from the interior zone into the silicate dust condensation zone, and
\begin{equation}
v_\mathrm{esc}^2={2GM_*\over r_0}
\end{equation}
is the escape velocity at distance $r_0$. At very large distance $r\gg r_0$ the velocity approaches 
\begin{equation}
v_\infty^2=v_0^2-v_\mathrm{esc}^2+{L_*\over 2\pi c r_0}\bar\varkappa_\mathrm{pr}\,,
\end{equation}
where
\begin{equation}
\bar{\varkappa}_\mathrm{pr}=\sum_i\kappa_{1,i}^{\mathrm{pr}}\psi_i\ \bar f_i\,,
\end{equation}
corresponds to the pressure extinction coefficient defined by Eq.~(\ref{DefKapPrMix})
calculated with an effective degree of condensation
\begin{equation}
\bar f_i=r_0\int_{r_0}^\infty {f_i\,\mathrm{d}r\over r^2} 
\end{equation}
of the different dust species.

The requirement that there is an accelerated outflow driven by radiation pressure on dust ($v_\infty>v_0$) requires
\begin{equation}
\Gamma_\mathrm{eff}={L_*\bar{\varkappa}_\mathrm{pr}\over4\pi cGM_*}\ge1\,,
\label{DefGammaAccel}
\end{equation}
which defines for given stellar mass and given mass-extinction of the matter the minimum luminosity required to drive the matter out of the gravitational potential trough of the star, the Eddington limit
\begin{equation}
L_\mathrm{edd}={4\pi cGM_*\over\bar{\kappa}_\mathrm{pr}}\,.
\label{DefEddLum}
\end{equation}
This limit luminosity is shown in Fig.~\ref{FigHR} separately for silicate dust with a Mg\# of 0.5 and for corundum dust for the case that either Si or Al are completely condensed into dust. The requirement $\Gamma_\mathrm{eff}>1$ is satisfied for all stars of our sample once silicate dust is formed, but not in the inner part of the outflow where only corundum dust is stable.  

With the assumption of a stationary, spherically symmetric outflow the  radial variation of the total mass density, $\varrho$, follows from
\begin{equation}
\dot M_\mathrm{g}=4\pi r^2 \varrho v\,,
\label{EqRhoVar}
\end{equation}
where $\dot M_\mathrm{g}$ is the (constant) rate of mass-loss. Multiplying Eq.~(\ref{EqVel}) with $4\pi r^2\varrho$ we obtain with Eq.~(\ref{EqRhoVar}) 
\begin{equation}
{\textrm{d}\,v\over\textrm{d}\,r}=\bar{\varkappa}_{\mathrm{pr}}\varrho\ 
{L_*\over c\dot M_\mathrm{g}}\ \left(1-{1\over\Gamma_\mathrm{eff}}\right)\,,
\label{EqVelAlt}
\end{equation}
which is an alternative version of the equation for $v$.
The stellar mass enters into calculations of the outflow velocity via $\Gamma_{\mathrm{eff}}$, which is usually significantly larger than unity, except at the inner edge of a dust shell at the onset of condensation. This means that the usually highly uncertain stellar mass has only as small impact of the velocity of the outflow.

\begin{figure}

\includegraphics[width=\hsize]{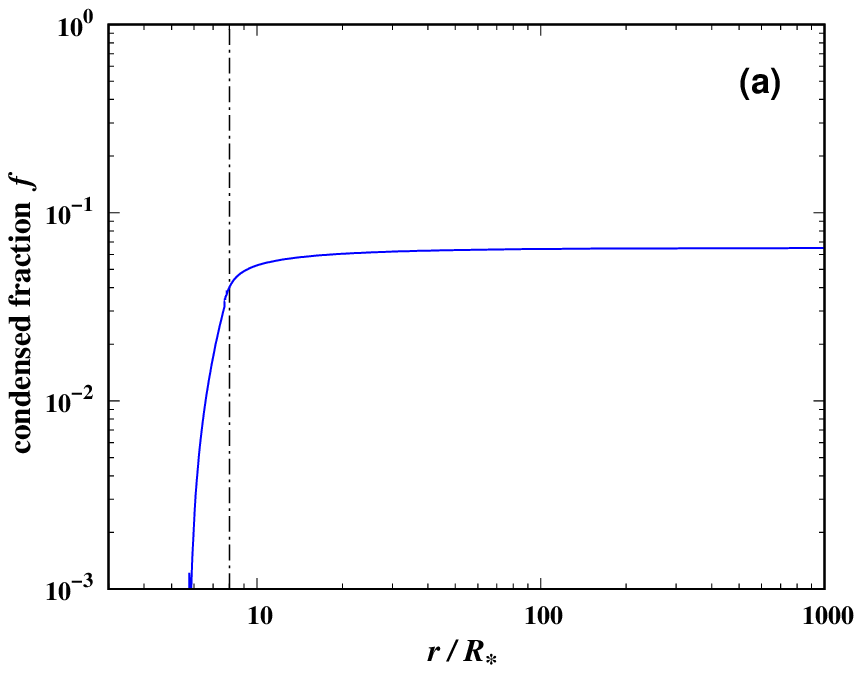}

\includegraphics[width=\hsize]{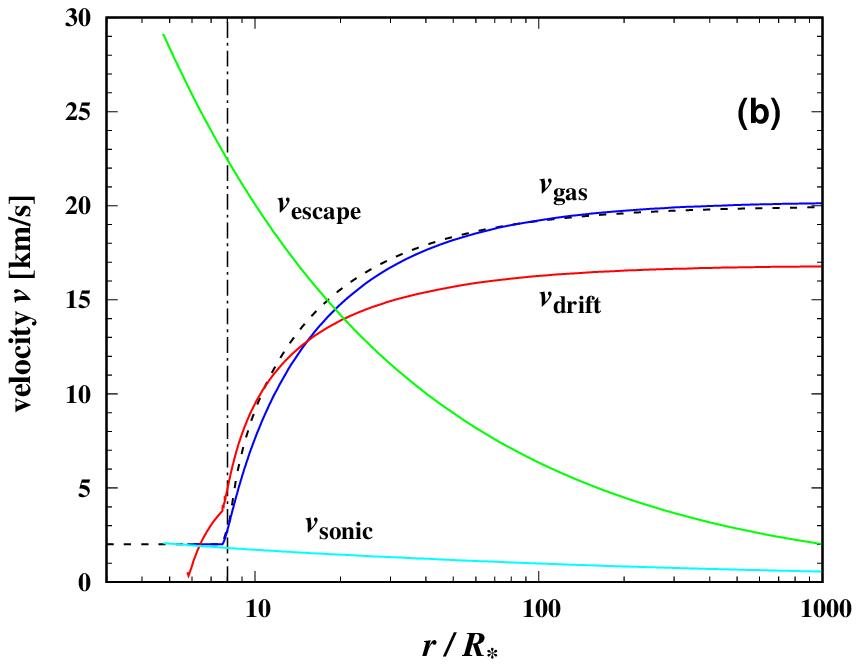}

\caption{Dust growth model for the silicate dust component for the supergiant $\mu$ Cep. (a) Radial variation of the fraction $f$ of the element Si condensed into silicate dust. (b) Outflow velocity of the gas, $v_\mathrm{gas}$, caused by radiation pressure on the silicate grains (solid blue line) and drift velocity of the silicate grains with respect to the gas, $v_\mathrm{drift}$, calculated for spherical grains in the tight coupling approximation (solid red line) and the simplified velocity law defined by Eq. (\ref{EqFlowSimpl}) (dashed black line). For comparison, also the variation of sonic velocity, $c_\mathrm{sonic}$, and escape velocity from the star, $v_\mathrm{excape}$, with radial distance are shown.}

\label{FigModGrowthMuCep} 
\end{figure}

\subsection{Instantaneous condensation}

Equation (\ref{EqVel}) can be integrated once the radial variation of $\varkappa_\mathrm{pr}$ is known. The mass-extinction coefficient $\varkappa_\mathrm{pr}$ varies (i) because the quantity of dust present in the gas-dust mixture (described by $f_i$) increases by dust growth and (ii) because the spectral composition of the radiation field varies. 

Here we assume that dust forms immediately once the temperature drops below a certain condensation temperature, that is, one either has no dust or a constant amount of dust per unit mass. This instantaneous condensation approach is frequently used for analysing the infrared emission from dust shells and is used as a first step in our model calculations. The second cause of $\varkappa_\mathrm{pr}$-variations does not apply in the case of optically thin shells (which are considered in this paper) because the spectral energy distribution is invariantly that of the stellar radiation field. 

With these assumptions $\varkappa_\mathrm{pr}$ is piecewise constant and only changes (idealised as a jump in the instantaneous condensation approximation) if a new dust species condenses. In particular we have in the present case
\begin{equation}
\bar f_i=f_i{r_0\over R_i}\,,
\end{equation}
where $R_i$ is the inner edge of the dust shell of species $i$

We have to integrate Eq.~(\ref{EqVel}) from the radius $R_k$ of that dust species $k$ for which $\Gamma_\mathrm{eff}>1$ for the first time if we move outwards. For our problem this species is always the silicate dust. For aluminium oxide dust we always have $\Gamma_\mathrm{eff}<1$ and the other possible components (iron, silicate with iron inclusions) appear at lower temperature. Since the contribution to $\varkappa_\mathrm{pr}$ of the components appearing at lower temperature is much lower than the contribution from the silicate dust, we neglect the fact that there is some minor increase in $\varkappa_\mathrm{pr}$ at some radii in between of $R_k$ and $R_\mathrm{a}$ and assume $\varkappa_\mathrm{pr}$ to be constant in the whole range with a value corresponding to the contribution of all species. 

In this approximation Eq.~(\ref{EqVel}) may be integrated with the result
\begin{equation}
v^2=v_0^2+(v_\infty^2-v_0^2)\left(1-{r_0\over r}\right)\,,
\label{EqFlowSimpl}
\end{equation}
where $r_0$ is the initial velocity at the inner edge $r_0$ of the dust shell (equal to the $R_k$ from above) and $v_\infty$ the final outflow velocity, or terminal velocity, at very large radius. This velocity, given by
\begin{equation}
v_\infty^2=v_0^2+{L_*\varkappa_\mathrm{pr}\over2\pi cr_0}\left(1-{1\over \Gamma_\mathrm{eff}}\right)\,,
\label{TermVel}
\end{equation}
corresponds to the outflow velocity of the stellar wind determined, for example, from observed molecular line widths of CO emission lines or of OH-masers. This velocity model was already used by \citet{Bed87} and \citet{ScT89} to model infrared emission from dust shells and found to enable good model fits. 

Figure \ref{FigModGrowthMuCep} compares the velocity model with the result of a model calculation for the coupled problem of wind acceleration and dust growth as described in Sect.~\ref{SectDuGrMod} for the special case of the supergiant $\mu$ Cep. Figure~\ref{FigModGrowthMuCep}a shows the radial variation of the degree of condensation and  Fig.~\ref{FigModGrowthMuCep}b the velocity increase due to dust condensation. The instantaneous condensation approximation obviously results in a reasonable approximation for the velocity variation of the dust-driven wind.

We wish to apply this approximation to stars, for which $v_\infty$ is known from observations. What we do not know is the initial velocity $v_0$ with which the outflowing gas enters the silicate condensation zone. We have to determine this as part of our fitting procedure by which we try to reproduce observed infrared spectra by model spectra.

The density structure of the dust shells, $\varrho_i(r)$, is given for each species in the instantaneous condensation approximation by an equation analogous to Eq.~(\ref{EqRhoVar}) 
\begin{equation}
\dot M_i=4\pi r^2 \varrho_i v
\label{EqRhoVarSpe}
\end{equation}
with constant mass-loss rates $\dot M_i$ of the dust species because we assume a stationary ouflow and no drift of the dust particles with respect to the gas. Observing this and Eq.~(\ref{EqRhoVar}), Eq.~(\ref{DefKapPrMix}) for $\varkappa_{\mathrm{pr}}$ can be written as
\begin{equation}
\varkappa_{\mathrm{pr}}={1\over\dot M_\mathrm{g}}\sum_i \kappa_{1,i}^{\mathrm{pr}}\,\dot M_i\,.
\label{DefKapPrMixInst}
\end{equation}
The mass-loss rates $\dot M_i$ are related to the mass of dust, $M_i$, contained in the shell between $R_i$ and $R_\mathrm{a}$ by
\begin{equation}
M_i=4\pi \int_{R_i}^{R_\mathrm{a}} r^2\varrho_i\,\mathrm{d}r=\dot M_i\,t_i^{\mathrm{exp}}\,,
\label{DefMpuDu}
\end{equation}
where
\begin{equation}
t_i^\mathrm{exp}=\int_{R_i}^{R_\mathrm{a}}{\mathrm{d}r\over v}
\label{DefTimeExp}
\end{equation}
is the time required for a gas shell to expand from $R_i$ to $R_\mathrm{a}$.

For given $\dot M_i$ the condensation fractions, $f_i$, can be calculated from
\begin{equation}
f_i={\dot M_i\over\psi_i\dot M_\mathrm{g}}\,.
\label{DefCondDeg}
\end{equation} 

\subsection{Application to dust models}
\label{SectCalcDotM}

In a more realistic approach, Eq. (\ref{EqVel}) has to be combined with an equation for the dust growth and both have to be solved numerically as it is discussed in Sect.~\ref{SectDuGrMod}. It is instructive, however, to study first the simpler approach just outlined. In this simple approach, expression (\ref{EqFlowSimpl}) can be used for the velocity variation in the outflow. This accounts at least qualitatively for the velocity increase in that part of the dust shell where most of its mid-infrared radiation is emitted. The density variation resulting from this is used for calculating the infrared emission from the dust shell. Fitting this to the observed mid-infrared spectrum of a star we obtain the dust masses $M_i$ and the radii $R_i$ at the inner edge of the corresponding dust shell for all dust species $i$. We also know the quantities  $\kappa_{1,i}^{\mathrm{pr}}$ which are calculated from optical constants. 

If we know from observations of molecular or atomic species the outflow velocity $v_\mathrm{exp}$ of the gas we can identify this with the terminal velocity $v_\infty$ and use this piece of additional information besides the infrared spectrum to obtain more information on the dust condensation process. This requires, however, that the motion of the dust grains is tightly coupled to the gas such that there is no substantial drift of the dust grains with respect to the gas and the outflow velocity of the dust is approximately the same as the observed outflow velocity $v_\mathrm{exp}$ of the gas. 

Then, letting $v_\infty$ in the above equations equal $v_\mathrm{exp}$ and assuming some value for $v_0$, one calculates
\begin{enumerate}

\item The timescales $t_i^\mathrm{exp}$ from Eq.~(\ref{DefTimeExp}).

\item The dust mass-loss rates $\dot M_i$ from Eq.~(\ref{DefMpuDu}).

\item The value of $\varkappa_\mathrm{pr}$ compatible with $v_\mathrm{exp}$ and the other pameters from Eq. (\ref{TermVel})

\item The gas mass-loss rate $\dot M_\mathrm{g}$ compatible with that value of $\varkappa_\mathrm{pr}$ from Eq.~(\ref{DefKapPrMixInst}).

\item The degrees of condensation $f_i$ from Eq.~(\ref{DefCondDeg}).

\item A new value for $\varkappa_\mathrm{pr}$ from Eq. (\ref{DefKapPrMix}).

\item A new value for the terminal velocity $v_\infty$ from Eq.~(\ref{TermVel}).

\end{enumerate}
Since the quantity $\Gamma_\mathrm{eff}$ in Eq. (\ref{TermVel}) already depends on the $f_i$, the equations have to be solved iteratively. The initial velocity $v_0$ then can be determined by varying $v_0$ such that the calculated value for $v_\infty$ obtained in step 7 equals the observed value~$v_\mathrm{exp}$.

One result of this procedure is the mass-loss rate $\dot M_\mathrm{g}$ for the gas. This may be compared to mass-loss rates determined from observations of atomic or molecular lines. We have chosen our sample of comparison stars such that for all of them this kind of information is available. This allows then a consistency check for our dust shell models. Another result are values for the $f_i$ which can be compared with results of solving the equation for grain growth  in the outflow.


\begin{figure}

\includegraphics[width=\hsize]{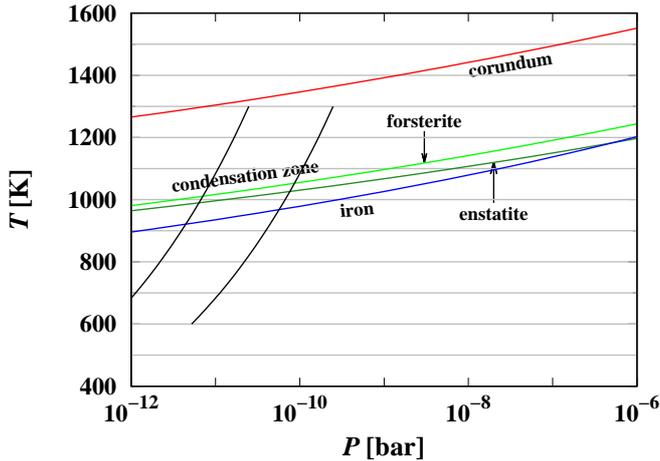}

\caption{Upper limit for thermal stability of the dust components in case of thermal equilibrium. The two black lines show the possible $p$-$T$ values at the inner edge of a dust shell for an assumed wind velocity of 2 km$^{-1}$ and mass-loss rates of $10^{-5}$ (right) and $10^{-6}\ \mathrm{M}_{\sun}\,a^{-1}$ (right) for a stationary outflow.}

\label{FigStabLim}
\end{figure}

\section{Comparison with observations}

\label{SectModl}

We consider in the following two types of models, the classical model with constant outflow speed that is frequently used to study dust properties in circumstellar shells, and the simple type of model where the outflow is driven to supersonic velocities by radiation pressure on dust.
    
\subsection{Constant velocity models}

In a first step we consider the model with constant outflow velocity as in paper I. We have a number of dust species, characterised by an index $i$, heated by the central star and re-emitting the absorbed energy in the infrared spectral region. It is assumed that the instantaneous condensation approximation can be applied where each of the dust species dust is present below a certain temperature $T_{\mathrm{a},i}$, characteristic for each of them, and no dust of this kind exists above $T_{\mathrm{a},i}$. This temperature, the appearance temperature, and the corresponding amount of dust present in the dust shell, $M_i$, is determined by comparing synthetic model spectra with the observed IR-spectra for the set of our comparison stars. The purpose is to get an overview which dust species are abundant and at which temperature they appear first in the outflow.

The model calculation consists of two parts, a calculation of the dust temperatures, $T_i(r)$, and a determination of the mass-densities, $\varrho_i(r)$,  of dust for the different dust species $i$. For optically thin shells the two parts are independent of each other.

The calculation of the dust temperature is done as described in paper I. A prescription of the temperature of appearance, $T_{\mathrm{a},i}$, for each dust species fixes the radius of the inner edge of the dust shell, $R_i$, of a dust species $i$.

In a stationary outflow, the density varies as 
\begin{equation}
\varrho_i(r)={\dot M_i\over 4\pi}\,{1\over r^2\,v_i}\,,
\end{equation}
where $\dot M_i$ is the dust mass-loss rate for species $i$ and $v_i$ the flow velocity of this dust species. The velocities $v_i$ equal the gas velocity, $v_\mathrm{g}$, if there is no (or at least no substantial) drift of dust grains with respect to the gas; otherwise they may be different for each species. The gas velocity may vary because of radiation pressure on dust and tight momentum coupling between gas and dust. Further, since the dust shells are very extended and the matter needs dozens of years to flow through the inner part of the shell which contributes the lion share of the mid-infrared emission, the mass-loss rate $\dot M_i$ may vary over that time. We follow custom practice and assume a density variation of the type
\begin{equation}
\varrho_i(r)=\varrho_i(R_i)\,R_i^p/r^p_{\,}\,.
\end{equation}
The power $p$ is considered as an unknown variable the value of which has to be determined from model fitting. For a stationary flow with constant outflow velocity one has $p=2$. A value $p\ne2$ means that the flow deviates from this simple case; this mimics a density stratification which may result, for instance, from a varying mass-loss rate. 

Then, with respect to the radial variation of the dust density we consider two types of models: 
either the exponent $p$ is set to a value of $p=2$, or the exponent $p$ is determined by optimisation. The purpose of the two models is to check if the dust shell can be described by the simple model of a constant outflow or if for any of our objects there are indications for a significant variation of the mass-loss rate during the recent past. 

The model assumes that the dust shell extends between the inner radius $R_i$ and a fixed outer radius $R_\mathrm{a}$ (common for all species). The prescription of the total mass $M_i$ of dust species~$i$ between $R_i$ and $R_\mathrm{a}$ uniquely determines $\varrho_i(r)$.

As dust species we consider the iron-bearing amorphous and non-stoichiometric silicates from paper I, a dirty silicate containing metallic iron as inclusions, amorphous corundum, and separate metallic iron grains as described in Sect.~\ref {SectAbsCo}. For the magnesium-iron silicate we have to prescribe for the model calculation also the composition, characterised by the Mg\#, and for the dirty silicate the volume fraction of metallic iron, $f_\mathrm{inc}$, included in the silicate ground mass.

For the size spectrum of the dust grains we assume a MRN distribution. The lower bound for the radius, $a_\mathrm{min}$, is set to $0.005\ \mu$m. The upper bound, $a_\mathrm{max}$, is generally set to $0.2\ \mu$m, but its value can also be included in the model optimisation process. Test calculations showed that treating $a_\mathrm{max}$ as a parameter to be determined by the model fit results in some cases in somewhat bigger radii (up to $\sim1\ \mu$m), but without leading to a definite improvement of the fit-quality. 

The dust opacity for the temperature determination and for the radiation pressure is calculated with Mie theory for spherical particles because the energy absorption and radiation pressure are determined by the stellar radiation in the optical to near infrared spectral region. In that wavelength range the condition $2\pi a\ll\lambda$ that the small-particle limit can be applied is not strictly satisfied, but the particles are still smaller than the wavelengths where the flux of the stellar radiation field peaks. For calculating the emission spectrum from the dust shell in the mid- to far-infrared spectral region, on the other hand, we used the CDE-approximation for small particles for the reasons discussed in Sect.~\ref{SectSizeShape}. In that spectral region the small particle approximation is valid. This procedure means a slight inconsistency in the model calculation which is accepted to avoid computationally more expensive methods to calculate for the optical to near infrared region the opacity of agglomerated particles. 

\subsection{Molecular bands}

\label{SectMolSph}

The observed spectra of supergiants frequently show in the mid-infrared spectral region a number of absorption bands from molecules residing in the stellar envelope. Most conspicuous are the strong dips in the energy distribution around 4.6 and 2.5 $\mu$m from CO (fundamental and first overtone), around 8 and 4.2 $\mu$m from SiO (fundamental and first overtone), and around 6.6 $\mu$m from H$_2$O. This absorption has its origin in a layer above the stellar photosphere located at about 1.5 $R_*$, the so called MOLsphere \citep{Tsu00,Tsu06,Per07,Ohn14,Mont14}. It is only observed for massive supergiants. 

From these molecular bands the SiO fundamental band long-wards of 8 $\mu$m falls on the short-ward flank of the 10 $\mu$m silicate feature and just blends the transition region between dominating silicate or iron absorption. This makes it difficult to judge how well the calculated infrared emission from the silicate band fits the observed infrared spectrum in that region. Therefore, we account, at least approximately, for this SiO absorption by multiplying the stellar spectrum with a correction factor
\begin{equation}
\Phi_\lambda={\rm e}^{-p_\lambda\cdot\tau_\mathrm{SiO}}\,,
\end{equation}
where the profile $p_\lambda$ is a smooth averaged absorption profile (normalised to unity at the wavelength of strongest absorption) and $\tau_\mathrm{SiO}$ the optical depth at that wavelength. A qualitative profile $p_\lambda$ was determined from \citet{Ver06} by constructing a smooth curve approximating the SiO profile given in their Fig.~3. The optical depth $\tau_\mathrm{SiO}$ is determined by treating this as an additional model parameter which is determined as part of the fitting procedure between observed and calculated spectrum.
 
For the other bands no correction of this kind is applied. The corresponding wavelength intervals can simply be excluded from the comparison of the spectra because no important information on the dust is contained in them. 

\begin{table}
\caption{Admitted range of parameters for optimisation.}

\begin{tabular}{llr@{\hspace{1mm}}c@{}ll}
\hline
\hline
\noalign{\smallskip}
Quantity & Variable & \multicolumn{3}{c}{Range} & Unit \\
\noalign{\smallskip}
\hline
\noalign{\smallskip}
Dust mass        & $M_i$              & 10$^{-10}$ &\dots&10$^{-4}$ & $M_{\sun}$ \\
Silicate         & $T_{\mathrm{a},i}$ & 300 &\dots& 1\,100 & K\\
                 & Mg\#               & 0.4 &\dots&  0.8 \\
Dirty Silicate   & $T_{\mathrm{a},i}$ & 300 &\dots&  900 & K \\
                 & $f_\mathrm{inc}$   & 0.0 &\dots&  0.5 \\
Corundum         & $T_{\mathrm{a},i}$ & 1\,200 &\dots&  1\,600 & K \\
Iron             & $T_{\mathrm{a},i}$ & 300 &\dots&  900 & K \\
Exponent density & $p$                & 1   &\dots&  3   \\
Optical depth SiO & $\tau_\mathrm{SiO}$ & 0 &\dots& 3 \\
Initial velocity & $v_0$          & 2   &\dots&  10 & km\,s$^{-1}$ \\
\noalign{\smallskip}
\hline
\end{tabular}

\label{TabRangeVar}
\end{table}

\subsection{Optimisation}

Synthetic spectra are calculated for a given set of model parameters as described in paper I. The parameters for each dust species are the quantities $M_i$ and $T_{\mathrm{a},i}$, for the magnesium-iron silicate the composition described by Mg\#, for the dirty silicate the volume fraction of iron inclusions, $f_\mathrm{inc}$, and $\tau_\mathrm{SiO}$. The stellar parameters are set to the values given in Table~\ref{TabStarParm}. The parameter values for the dust are optimised such that the deviation of the model spectra from observed ISO spectra becomes a minimum. This is done by searching for a minimum of
\begin{equation}
\chi^2=\sum_n\left({F^*_n-F^\mathrm{mod}_n\over F^*_n}\right)^2+ \left(v_\mathrm{exp}-v_\infty\over\sigma_v\right)^2\,, 
\label{DefChi2}
\end{equation}
where $F^*_n$ and $F^\mathrm{mod}_n$ are the stellar and model flux, respectively, at a set of selected wavelengths $\lambda_n$. For the optimisation we used 54 wavelength between 3 $\mu$m and 25 $\mu$m. In the second term, $v_\mathrm{exp}$ is the observed outflow velocity, $v_\infty$ the calculated terminal velocity, and $\sigma_v$ the error of the velocity determination. This second term is only considered for models where the value $v_0$ of the initial velocity is included in the optimisation process and the outflow velocity is calculated for such models (see Sect.~\ref{SectWind}). The minimum search is done by means of a genetic algorithm as described in paper I.
 
The range of allowed values of the parameters was limited to avoid that the optimisation algorithm tries to explore unphysical ranges of the parameter space. This is in particular important for the temperatures $T_{\mathrm{a},i}$ in order that all values tried during the course of the optimisation stay below the vaporisation temperature of a dust component and significantly above the dust temperature at the outer radius $R_\mathrm{a}$ of the shell. The admitted range of parameter values in our calculation is shown in Table \ref{TabRangeVar}. 

The adopted upper limits of $T_{\mathrm{a},i}$ correspond to the sublimation temperature of the dust species at the low pressures in the outflow. Figure~\ref{FigStabLim} shows the pressure dependence of the upper stability limits calculated for pure substances as described in \citet{Gai99}. The region between the two black lines in the figure corresponds to  the possible values of pressure and temperature at the inner boundary of a dust shell for the case that the wind enters the condensation zone with a velocity of 2 km s$^{-1}$ and for the two mass-loss rates $10^{-6}$ and $10^{-5}\ \mathrm{M}_{\sun}a^{-1}$ which encompass the values of $\dot M_\mathrm{g}$ for our stellar sample. The upper limit of $T_{\mathrm{a},i}$ for the silicate corresponds the stability limit for chemi-sputtering, for the dirty silicate and for the iron it corresponds to the sublimation temperature of iron. 

In the case of aluminium oxide the upper limit of $T_{\mathrm{a},i}$ corresponds to an arbitrarily assumed sublimation temperature; a precise value cannot be given since the likely pressure in its formation region \citep[around two stellar radii according to interferometric observations, for instance ][]{Kar13} is presently not known. The corresponding lower limit is set such that substantial amounts of corundum appear, if ever, before silicate dust (which is merely an assumption). In most cases, however, we set $T_{\mathrm{a},i}$ for corundum to a fixed value of 1\,400~K as in paper I because the properties of the model only weakly depend on this.
 
\begin{sidewaystable}

\caption{Quality of fit, $\chi^2$, for models with constant outflow velocitz and for wind models with different assumptions with respect to the initial velocity.}

\begin{tabular}{l@{\hspace{.9cm}}cccc@{\hspace{.9cm}}cccccc@{\hspace{.5cm}}cc}
\hline\hline
\noalign{\smallskip} 
             & \multicolumn{4}{c}{Power law variation $\varrho^{-p}$} & \multicolumn{6}{c}{Dust driven wind} & \multicolumn{2}{c}{Observed\phantom{MM}}\\
\noalign{\smallskip} 
             & $p=2$    & \multicolumn{2}{c}{$p$ optimised} & $a_\mathrm{max}$ optimised  &  \multicolumn{3}{c}{$v_0=2$ km\,s$^{-1}$} & \multicolumn{3}{c}{$v_0$ optimised} &  \\
Object       & $\chi^2$ & $\chi^2$   & $p$   & $\chi^2$ & $\chi^2$ & $v_\infty$ & $\dot M$ & $\chi^2$ & $v_0$ & $\dot M$   & $v_\infty$ & $\dot M$ \\
\noalign{\smallskip} 
             &          &            &       &     &     & km\,s$^{/1}$ & $M_{\sun}$a$^{-1}$ &   & km\,s$^{/1}$ & $M_{\sun}$a$^{-1}$ & km\,s$^{-1}$ & $M_{\sun}$a$^{-1}$\\
\noalign{\smallskip}
\hline
\noalign{\smallskip}
$\mu$ Cep    & 0.0269           & 0.0254           & 1.954  & 0.0214  & 0.0239           & 19.2 & $7.3\times10^{-6}$ & 0.0327           & 2.97 & $6.5\times10^{-6}$ & 20  & $5.0\times10^{-6}$ \\
RW Cyg       & 0.0839           & 0.0923           & 1.956  & 0.0655  & 0.0947           & 21.0 & $7.4\times10^{-6}$ & 0.0975           & 3.38 & $7.7\times10^{-6}$ & 23  & $3.2\times10^{-6}$ \\
W Per        & 0.0939           & 0.0943           & 1.974  & 0.0799  & 0.101\phantom{0} & 16.0 & $2.0\times10^{-6}$ & 0.113\phantom{0} & 3.79 & $2.3\times10^{-6}$ & 16  & $2.1\times10^{-6}$ \\
RS Per       & 0.0820           & 0.0792           & 1.891  & 0.0574  & 0.0777           & 19.3 & $3.4\times10^{-6}$ & 0.0944           & 2.16 & $3.0\times10^{-6}$ & 20  & $2.0\times10^{-6}$ \\
$\alpha$ Ori & 0.148\phantom{0} & 0.113\phantom{0} & 2.104  & 0.124   & 0.109\phantom{0} & 16.3 & $1.1\times10^{-7}$ & 0.118\phantom{0} & 2.08 & $1.3\times10^{-7}$ & 15  & $1.5\times10^{-6}$ \\
$\alpha$ Sco & 0.0361           & 0.0344           & 2.121  & 0.0354  & 0.0337           & 18.6 & $7.6\times10^{-8}$ & 0.0397           & 2.41 & $8.0\times10^{-8}$ & 17  & $1.0\times10^{-6}$ \\
S Per        & 0.101\phantom{0} & 0.109\phantom{0} & 1.948  & 0.100   & 0.168\phantom{0} & 22.0 & $6.9\times10^{-6}$ & 0.221\phantom{0} & 2.74 & $5.4\times10^{-6}$ & 20  & $7.5\times10^{-6}$ \\
PZ Cas       & 0.0787           & 0.139\phantom{0} & 2.007  & 0.0627  & 0.117\phantom{0} & 27.3 & $1.6\times10^{-5}$ & 0.117\phantom{0} & 14.3 & $2.6\times10^{-5}$ & 30  & $8.3\times10^{-6}$ \\
VX Sgr       & 0.0925           & 0.0688           & 1.947  & 0.0527  & 0.0709           & 25.2 & $3.7\times10^{-5}$ & 0.0857           & 3.22 & $3.1\times10^{-5}$ & 25  & $2.0\times10^{-5}$ \\
\noalign{\smallskip}
\hline
\end{tabular}

\label{TabCompFit}
\end{sidewaystable}

The optimisation was performed for 100 individuals for each generation and for 300 generations. A number of such optimisations (ten in most cases) was run for each object  with  different seeds for the random number generator to ascertain that the minimum found is reproduced in several runs. For some of the `best models' so obtained a final improvement was performed with a gradient method taking the output from the genetic algorithm as starting value, but generally, because of the high number of generations and individuals used, this final step does not result in an improvement of the model. 

\begin{figure*}

\includegraphics[width=0.33\hsize]{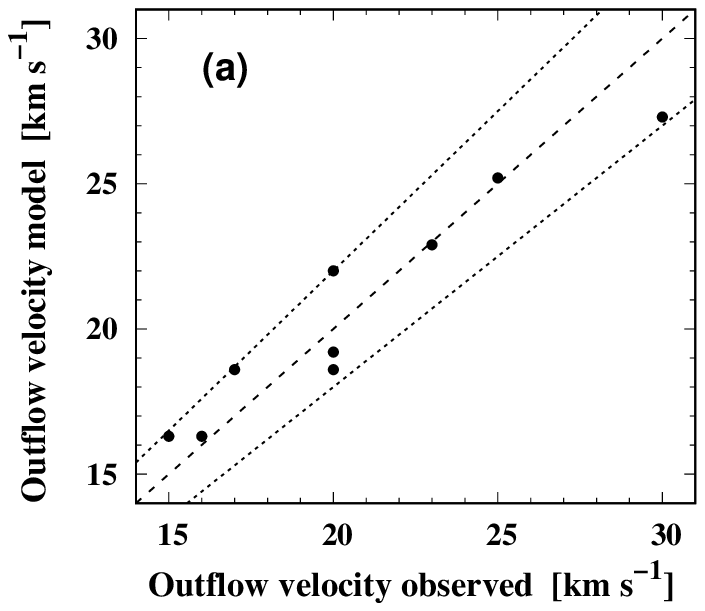}
\hfill
\includegraphics[width=0.33\hsize]{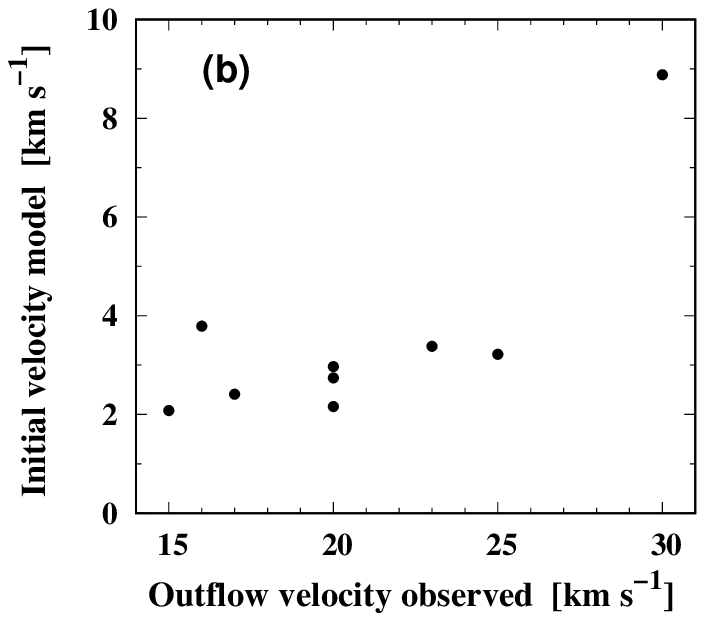}
\hfill
\includegraphics[width=0.33\hsize]{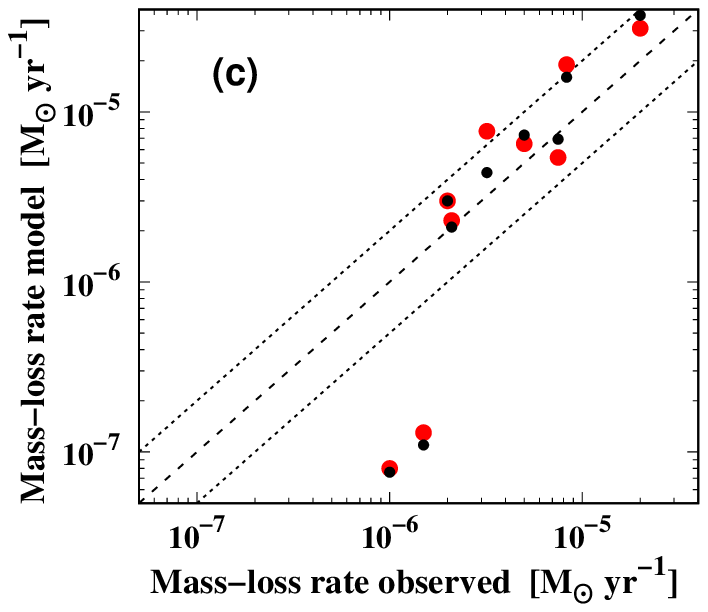}

\caption{Comparison of observed values with model results for the final outflow velocity, the initial velocity, and the mass-loss rate. (a) Terminal velocity for models with initial velocities $v_0=2$ km$s^{-1}$. The dotted lines show the limit of a deviation of the model result from the observed velocity by 10\%. (b) Initial velocity for models with given terminal velocities. (c) Mass-loss rates for models with initial velocities $v_0=2$ km\,s$^{-1}$ (black circles) and with given terminal velocities (red circles). The dotted lines show the limit of a deviation of the model result from the observed mass-loss rate by a factor of two.  }

\label{FigVelMpuFit}
\end{figure*}

For each of the comparison stars it was first attempted to determine an optimised model using all four dust components which we consider. It has to be observed in that case that not always the model with the lowest value of $\chi^2$ can be accepted because there are some limitations with respect to the real existence of the considered dust species which are not allowed for in the model calculation.   

First, there is some degeneracy with respect to dirty silicate and the metallic iron in the sense that for $f_\mathrm{inc}\gtrsim0.4$ the spectral variation of the absorption of the dirty silicate becomes undistinguishable from that of pure metallic iron (cf. Fig~\ref{FigIronIncl}). Therefore, if the optimisation process results in low values of $\chi^2$ for models where a high value of $f_\mathrm{inc}$ and comparable amounts of dirty silicate and metallic iron are predicted, we repeated the optimisation with omitting one of the two species and choose the mixture with the lower value of $\chi^2$. 

Second, the magnesium-iron silicate should appear at higher temperature than the dirty silicate because it is unlikely that there suddenly appears an inclusion-free silicate as a new component if temperature is already sufficiently low that dirty silicate is formed. Though the minimisation algorithm sometimes finds such models to represent a marked minimum of $\chi^2$ in parameter space, such models have to be rejected as unphysical. In all such cases it was found possible to find another optimised model which fits to the physical requirements. Since our optimisation problem is highly non-linear there are more than one local minima in the $\chi^2$-hypersurface, which do not all correspond to real physical situations.       

These ambiguities would not arise, if consistent growth models for all dust species were be implemented in the model calculation, which is out of the scope of the present paper. For the moment we eliminate unrealistic models from the results of the optimisation by inspection. 

\subsection{Optimised models for the stellar sample}

\subsubsection{Constant velocity models}

Table \ref{TabCompFit} compares the value of $\chi^2$ for the accepted models of the nine sample stars for the two kind of models where the parameter $p$ either is fixed to $p=2$ or is determined as part of the model optimisation. The best-fit models have comparable values of $\chi^2$ in both cases and the best-fit value for $p$ does not strongly deviate from $p=2$. In principle one would expect the models with optimised $p$ to result in a somewhat lower value of $\chi^2$ because one more parameter is used for fitting, but since the genetic algorithm operates with discrete values of the variables, this ideal case could only be approached with a very fine resolution of parameter space requiring a large computational effort. Here we content ourselves with the statement that for none of our objects a density structure significantly different from a stationary outflow is suggested by a much lower value of $\chi^2$. In the following we restrict our considerations to the case of stationary outflows.

Next we include the maximum size $a_\mathrm{max}$ of dust grains in Eq.~(\ref{DefDustKap}) of all four dust species in the optimisation process. We allowed this quantity to vary within the range from 0.1 to 10 times the value given in Table~\ref{TabCoSu}. The lowest value of $\chi^2$ that is obtained in this case is shown Table~\ref{TabCompFit}. For all objects the quality of the fit is somewhat improved. In most cases the new optimised maximum grain size is increased over our standard assumption, in particular for the alumina dust, but in some cases a reduced size yields a better fit. In all cases the reduction of the quality function by including $a_\mathrm{max}$ in the optimisation procedure is only slight. This probably results from the fact that in the mid-IR wavelength region used for spectral fitting the small-particle limit ($2\pi a\ll\lambda$) is almost valid, in which case the mass extinction coefficients do not strongly vary with size and shape of the grains. Since we do not obtain substantial better fits in this way we use in our further considerations only the fixed values of $a_\mathrm{max}$ given in Table~\ref{TabCoSu}.  

\subsubsection{Accelerated outflow models}

For the accelerated outflow model from Sect.~\ref{SectWind} we have to consider two different cases. Either the wind is due to the onset of silicate condensation and the associated strong radiation pressure on this dust species, or the wind originates closer to the star by some other process. 

In the first case the initial velocity $v_0$ to be used in Eq.~(\ref{EqWindVel}) has to be set to about the sound velocity at the inner edge of the silicate dust shell. We use in our calculation a typical value of $v_0=2$ km\,s$^{-1}$. As result one obtains, among other things, the terminal outflow velocity $v_\infty$ which can be compared to the observed value. 

In the second case the velocity at the inner edge of the silicate dust shell, $v_0$, is not known; it can take any value between the sonic velocity up to a large  fraction of the observed terminal outflow velocity. In this case we include $v_0$ as a parameter in the optimisation process and try to determine its likely value from the optimum fit of the model spectrum to the observed spectrum. For such models we have as additional model constraint that the calculated terminal outflow velocity should reproduce the observed value. This is included in the calculation of the quality function $\chi^2$ in the way shown in Eq.~(\ref{DefChi2}).  The resulting value of $v_0$ cannot be compared to an observed value, because no observational values seem to be known for this, but it can serve as a check for the consistency of the model used.

In both cases the mass-loss rate of the gas phase also results from the model fit (as described in Sect.~\ref{SectCalcDotM}). This can be compared to the corresponding value derived from observations of mass-loss for the gas phase.

Table~\ref{TabCompFit} shows the results for the velocity $v_0$ or $v_\infty$ and the gas mass-loss rates of the optimised fits for both types of models. Figure \ref{FigVelMpuFit} compares the observed mass-loss rates and terminal outflow velocities with the corresponding values obtained from the model fit. 

\begin{figure*}

\includegraphics[width=.32\hsize]{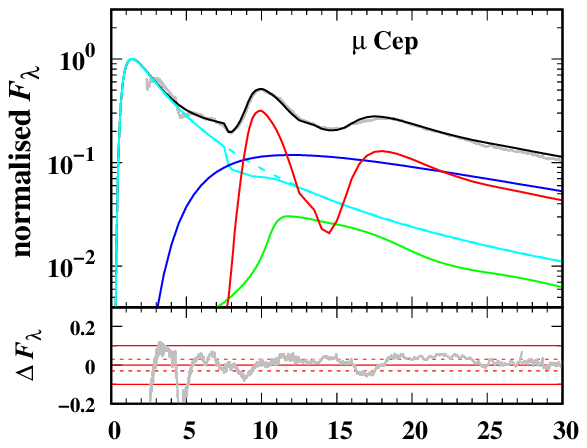}
\includegraphics[width=.32\hsize]{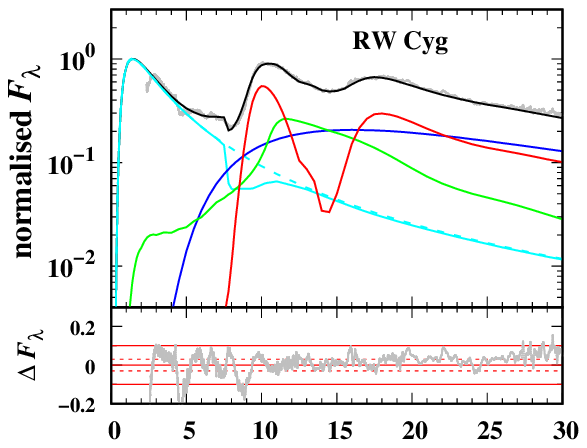}
\includegraphics[width=.32\hsize]{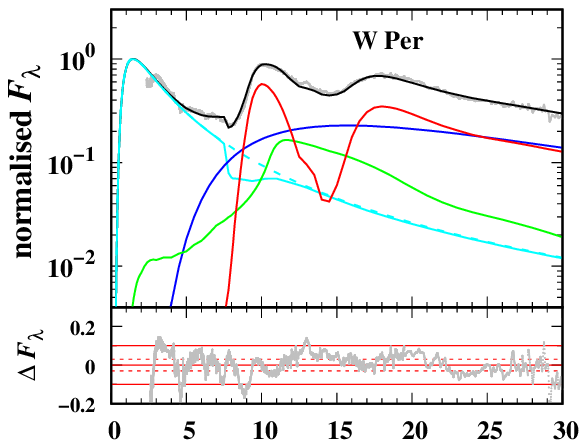}

\includegraphics[width=.32\hsize]{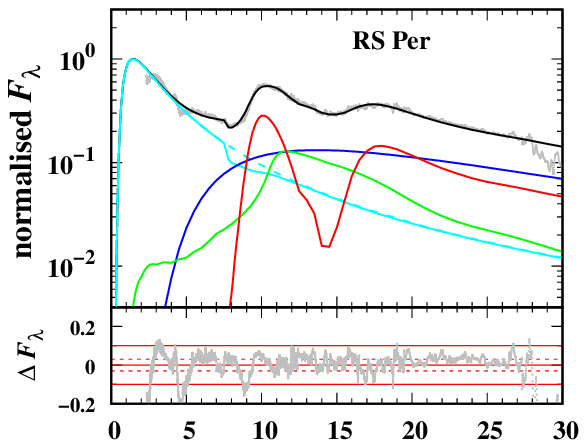}
\includegraphics[width=.32\hsize]{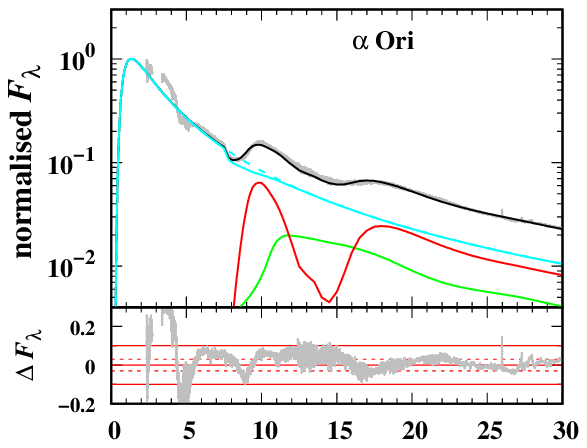}
\includegraphics[width=.32\hsize]{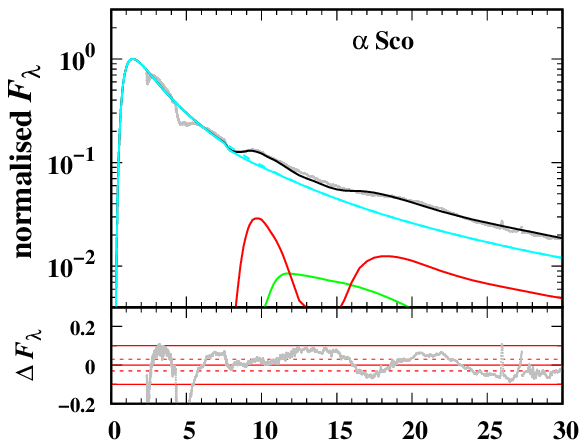}

\includegraphics[width=.32\hsize]{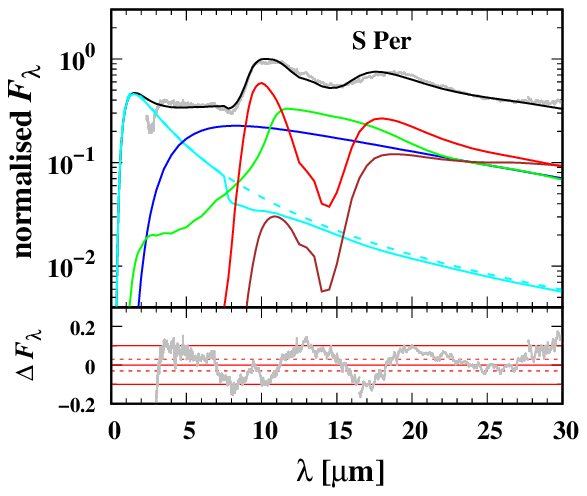}
\includegraphics[width=.32\hsize]{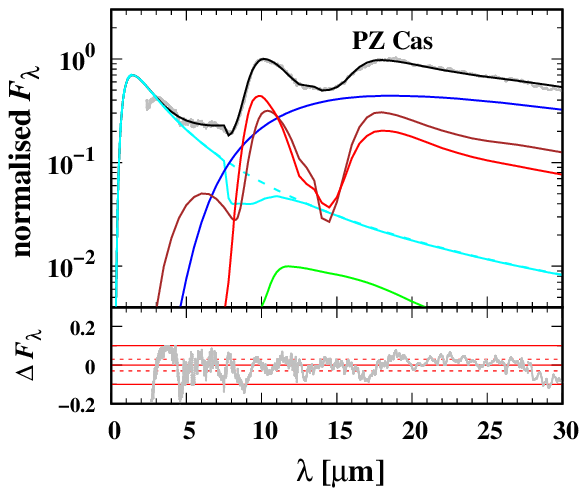}
\includegraphics[width=.32\hsize]{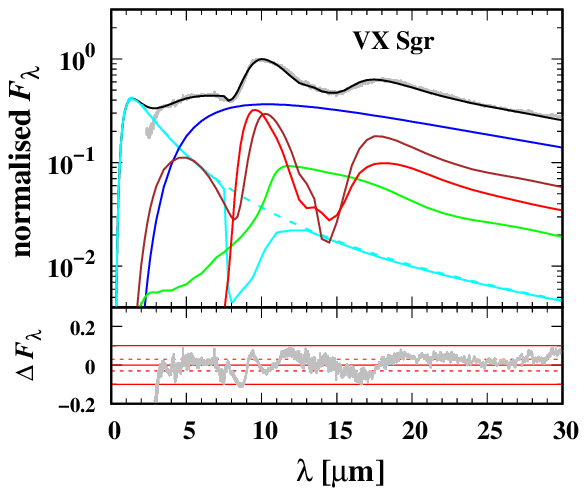}

\caption{Fit of ISO spectra of comparison stars (grey dots) with optimised synthetic spectra (solid black line). Individual contributions of the different dust components are: silicate material (solid red line), corundum (solid green line), iron (solid blue line), silicate with iron inclusions (solid dark-red line). The cyan line shows the stellar black body spectrum. The lower part of each picture shows the relative flux difference between observed spectrum and model. }

\label{FigSpecFit1}
\end{figure*}

\subsection{Comparison of velocity and mass-loss rate}

For the models with fixed initial velocity $v_0=2$ km\,s$^{-1}$ the resulting terminal velocity $v_\infty$ of the model agrees within $\pm11$\% accuracy with the observed values for all of the objects, see Fig.~\ref{FigVelMpuFit}a. In view of the many simplifications introduced in the outflow model used by us this agreement must be considered as rather good. This holds all the more since the observed terminal velocity is derived from Doppler widths of CO molecular lines which refer to a more distant region in the circumstellar shell than the region where the outflow is accelerated by radiation pressure on dust. This means, that the observed gas velocity refers to a state of stellar mass-loss at a different epoch (up to hundreds of years earlier) than the velocity calculated from the dust properties at the inner edge of the dust shell, and this could have changed in the meantime. But for none of our objects this seems to be an important effect, as we already found before.

For the models where the initial velocity at the inner edge of the silicate dust shell is varied and the final velocity forced to equal the observed outflow velocity, we find optimum models with a comparable fit quality to the case with fixed initial velocity. The admitted range of initial velocities was 2 \dots\ 15 km\,s$^{-1}$. In most cases the resulting value of the initial velocity of the optimum model is between 2 and 4 km\,s$^{-1}$, that is, close to our assumed value for the models with fixed initial velocity, see Fig.~\ref{FigVelMpuFit}b. The only exception is PZ Cas where a much higher initial velocity is obtained. In all cases the optimum value of the quality function $\chi^2$ is similar to that obtained for  models with given initial velocity. Hence, no convincing improvement is obtained by not to dispose of the initial velocity but by fixing the model by prescribing the terminal velocity instead. This holds even for the model for PZ Cas, where a still acceptable (but not optimal) fit can be found in this way.  

These model calculations show that it is possible to find a model for the dust shell that reproduces the infrared spectrum and the observed final gas outflow velocity with an outflow velocity at the inner edge of the silicate dust shell close to the sonic velocity ($\approx 1.9$ km\,s$^{-1}$ at 1\,000 K) or at least in the range of low supersonic velocities. A more precise value of $v_0$ at the inner edge of the silicate dust shell cannot be determined from our procedure.

Another check for the models is the gas mass-loss rate which can be determined once dust masses are derived, see Sect.~\ref{SectCalcDotM}. The results for $\dot M$ are shown in Table~\ref{TabCompFit}. These mass-loss rates are compared in Fig.~\ref{FigVelMpuFit}c to the gas mass-loss rates determined from atomic and molecular line emission. For six of our eight stars the gas mass-loss rates derived from the dust model are within a factor of two within the mass-loss rates derived from emissions from the gas phase. We cannot expect more than a rough agreement between observed and calculated rate, since most dust shells seem to have a clumpy dust distribution and since we have completely neglected the drift of dust relative to the gas. In view of the simple outflow model used by us and because no details of dust growth are considered, the results for $\dot M$ are considered in reasonable agreement between the dust model and gas phase observations. 

For two cases, $\alpha$ Ori and $\alpha$ Sco, the dust-derived mass loss-rates are by a factor of ten less than the mass-loss rates found from observations of molecular and atomic line emission. This significant discrepancy indicates, that the kind of outflow model considered by us is not applicable to these stars, possibly because of high clumpiness of the outflow, or significant variations of the mass-loss rate, or the UV-emission from a corona in case of $\alpha$ Ori \citep{Haa93}, or because of a deviating outflow geometry because of binarity in the case of $\alpha$ Sco (Sect.~\ref{SecBinAlfSco}). 

\begin{table*}

\caption{Properties of dust models with accelerated velocity and initial velocity $v_0=2$ km\,s$^{-1}$. The dust mass-loss rates are in units of $10^{-9}\,\rm M_{\sun}yr^{-1}$ and the temperatures in Kelvin. Species rejected by the optimisation are marked by `r'.}

\begin{tabular}{lcc@{\hspace{0.5cm}}rrrrrr@{\hspace{0.5cm}}rrrr@{\hspace{0.5cm}}rr}
\hline
\hline
\noalign{\smallskip}
     &   &  & \multicolumn{6}{c}{Mass-loss rate $\dot M_i$ and composition} & \multicolumn{4}{c}{Temperature $T_i$} \\[.1cm]
Object &  $\chi^2$ & $\tau_{10\mu\mathrm{m}}$  & Sil & Mg\# & Dirty & $f_\mathrm{inc}$ & Cor & Iron & Sil & Dirty & Cor & Iron & $\tau_\mathrm{SiO}$ & g/d \\
\noalign{\smallskip}
\hline
\noalign{\smallskip}
$\mu$ Cep    & 0.024 & 0.0140 &  2.47  & 0.61  &  r     &       & 0.190  & 0.309  &  910 &     & 1400 & 490 & 0.31 & 2451 \\
RW Cyg       & 0.095 & 0.0361 &  4.18  & 0.48  &  r     &       & 0.623  & 1.055  &  938 &     & 1400 & 335 & 0.99 & 1202  \\
W Per        & 0.101 & 0.0337 &  2.87  & 0.56  &  r     &       & 0.260  & 0.399  &  614 &     & 1400 & 343 & 0.60 & 631  \\
RS Per       & 0.077 & 0.0192 &  1.92  & 0.47  &  r     &       & 0.326  & 0.363  &  923 &     & 1400 & 424 & 0.33 & 1302 \\
$\alpha$ Ori & 0.109 & 0.0035 & 0.125  & 0.62  &  r     &       & 0.0324 &  r     &  981 &     & 1400 &     & 0.21 & 691  \\
$\alpha$ Sco & 0.034 & 0.0017 & 0.0886 & 0.80  &  r     &       & 0.0153 & 0.0044 &  816 &     & 1400 & 741 & 0.10 & 705  \\
S Per        & 0.168 & 0.0558 &  2.82  & 0.77  &  1.82  & 0.278 & 0.909  & 0.101  &  854 & 608 & 1400 & 432 & 2.41 & 1050 \\
PZ Cas       & 0.117 & 0.0612 &  51.2  & 0.63  &  4.63  & 0.372 & 0.683  & 0.706  &  402 & 400 & 1400 & 307 & 1.98 & 451  \\
VX Sgr       & 0.071 & 0.0484 &  7.82  & 0.79  &  5.62  & 0.249 & 1.021  & 1.905  &  827 & 606 & 1400 & 691 & 2.83 & 2291 \\
\noalign{\smallskip}
\hline
\end{tabular}

\label{TabDustModl}
\end{table*}

\subsection{Comparison of synthetic and observed spectra}

Figure \ref{FigSpecFit1} compares for the set of models with initial velocity $v_0=2$ km\,s$^{-1}$ the synthetic spectra of the best-fit models with observed ISO-spectra from the catalogue of \citet{Slo03}. The upper part of each picture shows the model spectrum as black solid line, normalised with the maximum value of the flux. The infrared spectrum as observed with the ISO satellite is shown with grey dots; this is vertically shifted in the diagram such as to match with the model spectrum at a wavelength of 7 $\mu$m where there are no prominent molecular absorption bands. For comparison, also the pure stellar spectrum (black body radiation assumed) is shown as solid line in cyan colour, and the effect of the absorption by the MOLsphere on the radiation (see Sect. \ref{SectMolSph}) is shown by the dashed line in cyan colour. The lower part of each picture shows the relative deviation $(F_\lambda^\mathrm{obs}-F_\lambda^{mod})/F_\lambda^\mathrm{mod}$ between the observed (shifted) and calculated flux. 

For most of the objects a set of model parameters is found for which outside of the regions of the strong molecular bands (in particular that of H$_2$O around 3 $\mu$m and 6.7 $\mu$m, CO at 4.6 $\mu$m, and SiO at 8 $\mu$m) the model spectrum reproduces the observed infrared spectrum within error bounds of $\pm10$\%. The few dust species used in our model suffice to reproduce the main absorption properties of the dust mixture formed in these objects. The remaining structure that can be seen in the residual spectrum shown in the lower parts of the pictures indicate that additional minor dust species are likely present. They are not considered here. 

Of special interest for us is the spectral range between $9.5\ \mu\mathrm{m}\lesssim\lambda\lesssim20\ \mu\mathrm{m}$ where the two strong silicate bands are located, and the range $5\ \mu\mathrm{m}\lesssim\lambda\lesssim8\ \mu\mathrm{m}$ where a contribution of metallic dust grains is expected. The spectral energy distribution of the dust emission in the range $9.5\ \mu\mathrm{m}\lesssim\lambda\lesssim20\ \mu\mathrm{m}$ can be reproduced rather accurately by using the non-stoichiometric iron-bearing silicates from paper~I, augmented by amorphous alumina and metallic iron for filling-up the absorption gap between the two strong silicate bands. The individual contributions of the different species to the total emission are shown as solid red line for the silicate component, as solid blue line for the iron component, and as solid green line for the alumina component. The close fitting of the observed spectral features of the silicate dust suggests that the structure of the silicate dust formed in the dust shell is not a mixture of olivine-like material with isolated SiO$_4$ tetrahedra and pyroxene-like material with chains of polymerised SiO$_4$ tetrahedra with two bridging $\dots\mathrm{Si-O-Si}\dots$ groups per tetrahedron, as it is often assumed in models of circumstellar dust shells, but more likely a material with a non-negligible fraction of bridging oxygen atoms between SiO$_4$ tetrahedra. This is in line with laboratory findings on the (Mg+Fe)/Si ratios in pre-solar silicate dust grains from evolved stars, which, to a large extent, take values between the limit cases of 1 and 2 for chain- and island-silicates, respectively \citep[cf. the review by][]{Flo16}.

The significant infrared excess emission in the range $5\ \mu\mathrm{m}\lesssim\lambda\lesssim8\ \mu\mathrm{m}$ over the emission from the stellar photosphere in all our models can be reproduced by assuming the presence of metallic iron dust grains, either as inclusions in a silicate matrix (the dark-red solid line in Fig.~\ref{FigSpecFit1}) or as separate metallic grains (the solid blue line in Fig.~\ref{FigSpecFit1}). The only exceptions are $\alpha$ Ori and $\alpha$ Sco, which require no or almost no such dust to be present.  

A presence of a dirty silicate component with iron inclusions besides a pure iron component is only found in the cases of S Per, PZ Cas, and VX Sgr to result in an improved fit to the spectral emission profile over a fit with iron alone. In all three cases the volume fraction of the inclusions is high such that the opacity of the composed material is strongly modified compared to the silicate ground mass. Such kind of material with iron volume fractions of about 0.35 or higher where the absorption properties of the mixture are ruled by the iron component are found as possible dust component in some model fits of the other six objects. These models were rejected because this kind of composed material and pure iron are not clearly discernable on the basis of their absorption properties and models including pure iron grains alone result in an as good fit quality as models including dirty silicates and iron. The `iron dust' in the first six models in Table~\ref{TabDustModl} may then well represent a mixed material with a high fraction of iron inclusions. In paper I it was argued that such kind of iron-doped silicates are detected as pre-solar grains in meteorites. 

The simple type of modelling a silicate component with iron inclusions by assuming either no such dust or a material with the full final content of inclusions is probably not adequate. A more realistic treatment that accounts for the sequential formation of first an inclusion-free silicate at higher temperatures, the development of a mixed material during subsequent cooling, and possibly formation of a final iron coating at low temperatures is presently not available.
  
All models require for a good fit to the observed spectrum the presence of alumina dust (the solid  green line in Fig.~\ref{FigSpecFit1}). Because of the high bond energy of this solid it can exist at higher temperature and much closer to the star than the silicate and iron dust. First, a possible range of appearance temperatures between 1\,200 K and 1\,600 K was admitted for the parameter optimisation process, but it turned out that the fit quality is not much better than in the case that a fixed value from this range is prescribed. The final optimisation then was done by assuming a value of 1\,400 K for the appearance temperature of alumina. The dust temperature of the inner edge of the alumina dust shell cannot precisely be fixed in our models. The reason seems to be the circumstance that the most important contribution to the spectrum by alumina dust is in the wavelength range corresponding to the gap between the two silicate bands. For this range emission by cooler dust from more distant layers is more important than that from the hot inner edge of the alumina dust shell. The value of 1\,400 K given in Table~\ref{TabDustModl} only indicates that alumina dust forms much closer to the star than silicate dust, as it seems to be also observed in interferometric observations \citep[for instance ][]{Dan94,Kar13}. 

The approach used by us for accounting for the molecular SiO absorption feature due to a MOLsphere (Sect.~\ref{SectMolSph}) does not completely remove the absorption band at $\sim8\ \mu$m seen in nearly all spectra of the supergiants. Further \citet{Ver09} found indications for the presence of PAH features which may also contribute to extinction around 8.8 $\mu$m. The treatment of this feature has to be done more carefully \citep[as for instance in][]{Ver06} if the details of the silicate $9.7\ \mu$m feature have to be compared in more detail with observations, but this is out of the scope of our paper. 

For all models the optical depth at 10 $\mu$m for the optimum fit is smaller than 0.1 (Table \ref{TabDustModl}). This justifies our approach to calculate the mission spectrum under the assumption that the shells are optically thin.

\begin{table}
\caption{Fraction of key elements condensed into dust for the optimum fit models with accelerated velocity and initial velocity $v_0=2$ km\,s$^{-1}$. }

\begin{tabular}{lccccc}
\hline
\hline
\noalign{\smallskip}
Object       & Sil & \multicolumn{2}{c}{Dirty}  & Cor & Iron \\
             & $f_\mathrm{Si}$ & $f_\mathrm{Si}$ & $f_\mathrm{Fe}$ & $f_\mathrm{Al}$ & $f_\mathrm{Fe}$ \\
\noalign{\smallskip}
\hline
\noalign{\smallskip}
$\mu$ Cep    & 0.0840 &        &        & 0.111 & 0.0321 \\
RW Cyg       & 0.147  &        &        & 0.375 & 0.113  \\
W Per        & 0.319  &        &        & 0.496 & 0.136  \\
RS Per       & 0.140  &        &        & 0.406 & 0.0807 \\
$\alpha$ Ori & 0.253  &        &        & 2.63  &        \\
$\alpha$ Sco & 0.274  &        &        & 1.65  & 0.046  \\
S Per        & 0.101  & 0.470  & 0.133  & 0.561 & 0.111  \\
PZ Cas       & 0.491  & 0.271  & 0.103  & 0.112 & 0.0207 \\
VX Sgr       & 0.516  & 0.282  & 0.072  & 0.116 & 0.0038 \\
\noalign{\smallskip}
\hline
\end{tabular}

\label{TabModFcond1}
\end{table}

\subsection{Condensation of dust}

Table \ref{TabDustModl} shows the resulting optimum-fit parameters for dust mass-loss rate and the corresponding condensation temperature for the nine stars from Table~\ref{TabStarParm}. Table~\ref{TabModFcond1} shows the fraction of the key elements of the different dust species which is condensed into dust.

We remark at this place that our best fit models are not strongly favoured in the sense that no other combination of model parameters result in a comparably good fit. For some of our objects models with a similar low value of the quality function but different values of some of the model parameters can be obtained. Such ambiguities concern mainly the value of the outflow velocity and the appearance temperature. The models presented here are that which better reproduce the observed final velocity or which have no implausible low appearance temperature of some dust species. In this sense the models may be subject to some bias.

\subsubsection{Degree of condensation}

The silicate features in all cases are reproduced the best with an iron-bearing amorphous, non-stoichiometric silicate with roughly equal abundances of Mg and Fe cations as the dominating dust species. Only for $\alpha$ Sco, S Per, and VX Sgr a more Mg-rich composition is required. The composition probably indicates that most of the condensed silicate material forms under conditions far below the stability limit for evaporation, where re-evaporation of Mg or Fe atoms from the growing particle after their attachment is negligible (= deposition on a cold substrate). In that case there is no preferential removal of the less stable iron component of the solid solution as it would be observed close to the stability limit under (near) equilibrium conditions. Since for present-day supergiants the element abundances of Mg and Fe are almost equal to each other (cf. Table~\ref{TabElAbu}) such that collisions with Mg and Fe atoms from the gas phase are almost equally frequent, one expects a value of $\approx0.5$ for Mg\# as it is found in our models for most of the objects.

\begin{figure}

\includegraphics[width=.8\hsize]{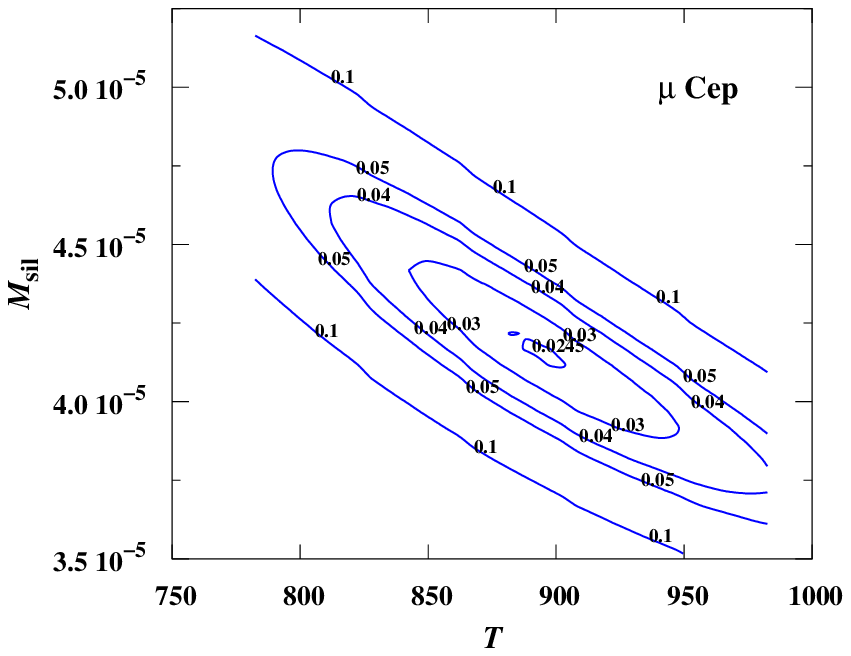}

\includegraphics[width=.8\hsize]{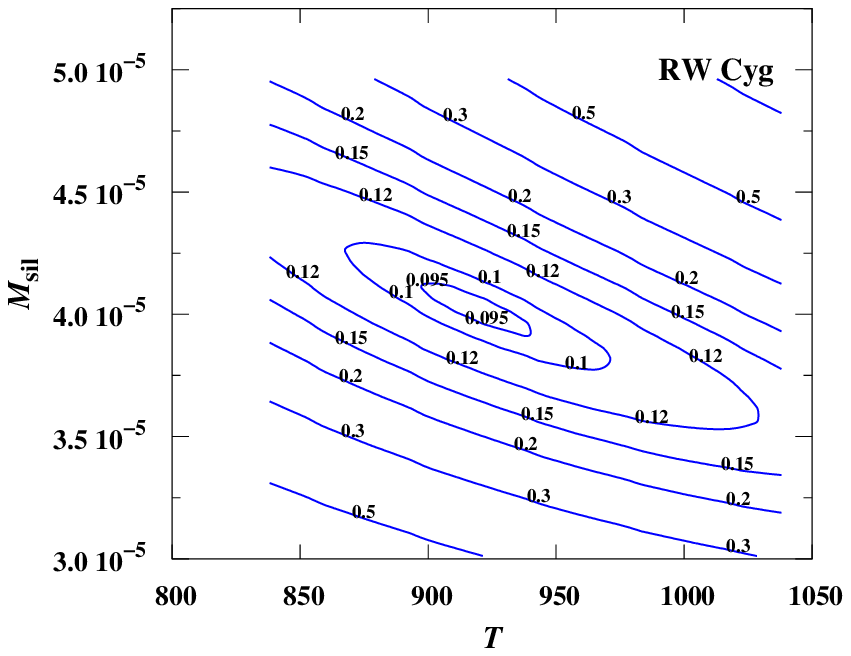}

\includegraphics[width=.8\hsize]{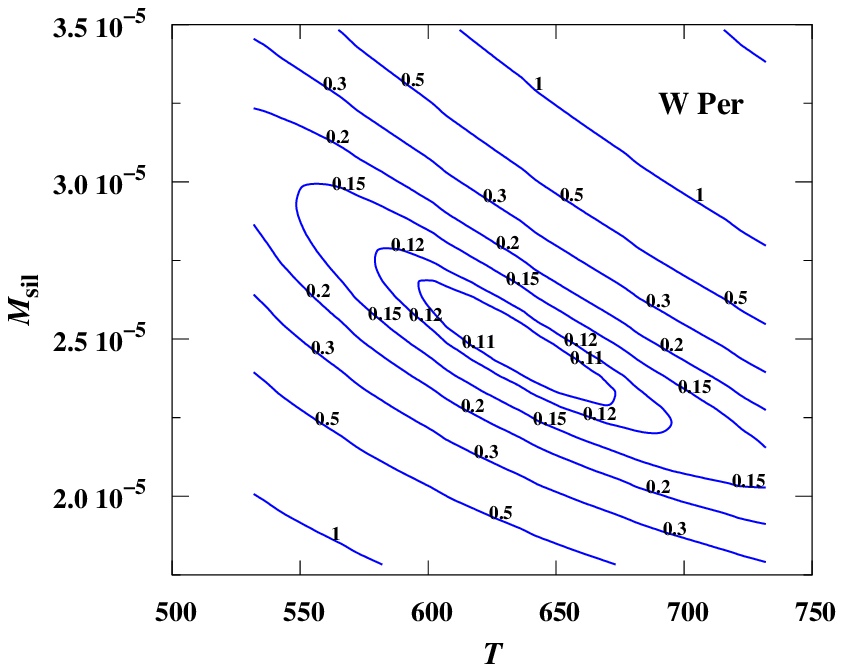}

\includegraphics[width=.8\hsize]{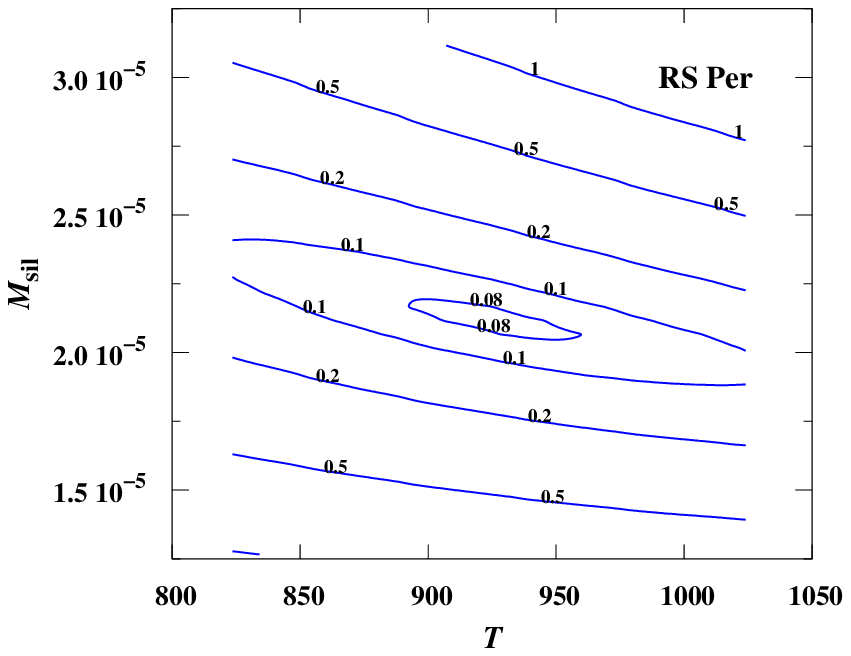}

\caption{Contourlines for $\chi^2$ for varying silicate mass, $M_\mathrm{sil}$, in the dust shell between $R_\mathrm{i}$ and $R_\mathrm{a}$, and varying appearance temperature $T_\mathrm{sil}$. All other parameters are set to their value for the optimum fit model. 
}

\label{FigVarChiMuCep}
\end{figure}

The fraction of Si condensed into silicate dust is shown in Table \ref{TabModFcond1}. For S Per, PZ Cas, and VX Sgr the combined fraction of the Si condensed into silicate dust with and without iron particle inclusion is 0.57 to 0.80. This high degree of condensation is difficult to explain with a growth model for dust and the simple accelerated outflow model used to derive the condensation fractions (see Sect. \ref{SectDuGrMod}). Though it is possible to obtain a reasonable fit between synthetic and observed spectra, the special outflow model used here is not applicable to these stars. We do not consider them further. 

For all objects save for $\alpha$ Ori an optimum fit requires the presence of metallic iron dust particles, either as nano-particle inclusions in silicate dust or as a separate dust component. Since the absorption of dirty silicate dust does not substantially differ from that of a mixture of silicate grains without iron inclusions and separate iron grains, it is difficult to discriminate between the two cases by spectral fitting alone. Only in the case of S Per, PZ Cas. and VX Sgr the best fit is obtained by including dirty silicate dust as a component of the dust mixture. In all other cases the optimum fit requires either a dirty silicate dust component with a high volume fraction of iron inclusion or a component of pure iron grains, but with negligible contributions by the other one of the two. Each of the two cases results in nearly equal values of the quality function $\chi^2$ at minimum. 

In optimum fit models with dirty silicates the resulting volume fractions of the iron inclusions are high, usually $f_\mathrm{inc}\gtrsim0.35$, in which case the extinction properties of the composite material are already dominated by the iron component (see Fig.~\ref{FigIronIncl}). This ambiguity could result from our assumption of a radially constant fraction of iron inclusions in the dirty silicate which in reality could well increase with increasing distance, for instance because of agglomeration of silicate and iron dust grains in the outflow, or by changing surface growth conditions with decreasing temperature. Such complexities are not considered in our model. This could result in a fit where in the range where the fraction of iron inclusions is low this dust is counted as inclusion-free silicate dust and in the range where the fraction of iron inclusions is higher this dust is counted as a mixture of inclusion-free silicate dust grains with either pure iron dust grains or with silicate dust grains with a very high fraction of inclusions.  

Because of this ambiguity we present for the objects  $\mu$ Cep, RW Cyg, W Per, RS~Per, and $\alpha$ Sco in Tables~\ref{TabDustModl} and \ref{TabModFcond1} only data for the models where no dirty silicate dust is included in the model. It is to be noted that in these models the iron dust could mask some additional portions of Si condensed into a composite iron-silicate dust material. It is also to be kept in mind that the source of the continuous extinction ascribed to iron instead could in part or totally be due to continuum emission from a warm chromosphere, as mentioned in Sect.~\ref{SectDustComp}, which introduces a further uncertainty of the interpretation. We plan to discuss the problem of the iron dust component in detail in a separate paper. 

The fraction of Al condensed into alumina is found to be rather high. For $\alpha$ Ori and $\alpha$ Sco it even exceeds unity, which contradicts its definition. For these two objects the type of model considered here is not applicable for this reason. Since the dust emission can well be fitted by a constant outflow velocity model (cf. Table~\ref{TabCompFit}) their wind is more likely driven by a mechanism operating close to the star \citep[cf.][]{Har96,Cro10} and dust forms passively in a strongly inhomogeneous and already supersonic outflow \citep[see for instance ][ for a discussion of the complex structure of the chromosphere and wind]{Har09,Lob00}. For this reason we do not further consider $\alpha$ Ori and $\alpha$ Sco, though $\alpha$ Sco is a particular intersting object because it is the only object for which depletions of dust forming elements in its outflow can be determined by absorption line spectroscopy \citep{Sno87,Baa07} since it is member of a binary with a hot O star companion (see Appendix~\ref{SecBinAlfSco}). 

For the objects $\mu$ Cep, RW Cyg, W Per, and RS~Per our model results seem to indicate that the alumina dust forms in a region inside of the silicate dust condensation zone where the average outflow velocity is low, probably subsonic. Because the atmospheres of late type supergiants are characterised by the presence of giant convection cells \citep[for instance ][]{Jos07,Chi11} there exist strong convective local in- and outflow motions in the photosphere and in the overlaying outer layers which manifest themselves in the strong observed broadening and deformation of spectral lines in their spectra \citep{Gra08}. The circumstances under which alumina dust is formed, thence are likely different from those which are generally assumed for circumstellar dust of AGB stars.

\subsubsection{Silicate condensation temperature}

An important information with respect to the question how the observed silicate dust forms in the outflow is the temperature for the onset of massive dust precipitation. Exactly this quantity cannot be derived from the type of models considered by us because dust condensation is not part of the model calculation. However, we can take the temperature at the onset of strong wind acceleration (our appearance temperature) as an approximation for this since already for an only small fraction $f\approx0.05$ of the Si condensed into amorphous iron-bearing silicate dust the quantity $\Gamma_\mathrm{eff}$, defined by Eq.~(\ref{DefGammaAccel}), exceeds unity and from that point on radiation pressure drives the dusty gas to the highly supersonic expansion velocity. Most of the material finally condensed into dust is condensed only afterwards. In this sense the  temperatures $T_i$ in Table~\ref{TabDustModl} indicate the temperature below of which most of the final dust is formed. 

For most of the objects this temperature is found to be in the range between somewhat above 900 K to somewhat below 900 K. Exceptions are W Per and PZ Cas for which the optimum-fit models require a significant lower appearance temperature.    

For the four objects to which the accelerated outflow model can be applied, Fig.~\ref {FigVarChiMuCep} shows contourlines of the quality function $\chi^2$ for models where the mass in the silicate dust shell between the inner edge $R_\mathrm{i}$ and the outer edge $R_\mathrm{a}$ and the appearance temperature $T$ of the silicate dust is varied around their values for the optimum-fit model and all other model parameters are set to their values for the optimum-fit model. As is seen, the iso-$\chi^2$ lines enclose strongly elongated and approximately elliptically shaped regions inside of which a model spectrum can be obtained that fits the observed spectrum with a value of the quality function better than a given value of $\chi^2$. From an inspection of the figure we estimate the temperature range within which the appearance temperature can be varied around the optimum-fit value without a significant loss of fit-quality to be about $\pm50$ K. This corresponds to a relative accuracy of $\sim\pm6$\%. At the same time, the amount of dust can be varied by about $\pm 0.25\times10^{-5}$ $M_{\sun}$ for $\mu$ Per and RW Cyg, and by $\pm 0.2\times10^{-5}$ $M_{\sun}$ for W Per and RS Per, around the optimum fit value. This corresponds to relative accuracies of about $\pm6$\% to $\pm8$\%. 

For the range of mass-loss rates between 10$^{-6}$ and $10^{-5}$ $M_{\sun}$a$^{-1}$ a condensation temperature around 900 K is approximately 100 K below the vaporisation limit of the crystalline silicate minerals forsterite and enstatite (see Fig.~\ref{FigStabLim}). The silicate dust formed in the outflow has an amorphous structure, and amorphous materials are generally thermally less stable than the corresponding crystalline materials. Though no data are available for the vaporisation of amorphous silicates with olivine or pyroxene structure, their vaporisation temperature must be lower than for the crystalline materials. The best-fit appearance temperatures of silicate dust as determined for $\mu$ Cep, RW Cyg, and RS Per mean that  silicate dust starts to be formed in significant quantities more or less immediately after the temperature dropped below the vaporisation limit of the material. 

Such a behaviour is generally observed if condensation happens by precipitation on pre-formed seed nuclei, while if initiation of condensation requires the nucleation of suited seed particles from the gas phase this is generally observed only at substantial supercooling below the stability limit. In the outflows from the late type supergiants, suitable pre-formed seeds for the precipitation of silicate dust are available in form of alumina dust grains. The presence of such grains is found to be required to explain the infrared emission in our models and also in the models of \citet{Ver09}. Our findings, thus suggest that for $\mu$ Cep, RW Cyg, and RS Per the silicate dust formed by growth on top of pre-formed seed nuclei. The growth on silicate dust on alumina grains is considered in detail in the next section. Alumina is chosen because this seems to be an abundant dust species and is highly refractory, but other kind of seed particles from refractory minerals \citep[eg. Ti-bearing compounds as proposed by][]{Gai98,Jeo99} presently cannot be excluded.

In case of W Per the optimum fit model requires a significant lower appearance temperature of 614 K, which even in view of the only moderate accuracy with which this temperature can be fixed by spectral fitting, is definitely much lower than the vaporisation temperature of the silicate dust. This may have two reasons. Either, the mass-loss rate in the past was significantly higher than presently such that the emission from the inner part of the dust shell is veiled by the dominant emission from cooler and earlier ejected material in more distant parts of the dust shell. Alternatively, the number density of alumina grains entering the inner edge of the dust shell is so small that only small quantities of silicate dust are formed by precipitation on alumina particles and that at lower temperature efficient nucleation from the gas phase commences which triggers copious silicate dust formation at low temperatures. How nucleation of silicates from the gas phase may work is still an open question. If it is initiated by clustering of the abundant SiO molecules, it would operate at temperatures around 630 K for mass-loss rates below $10^{-5}$ $M_{\sun}$a$^{-1}$, as discussed in \citet{Gai13a} (observe that the agreement with the temperature found for W Per must be considered as fortuitously). Nucleation involving MgSiO$_2$ or Mg$_2$SiO$_4$ molecules is proposed by \citet{Gou12,Gob16,Mau18,Val20}, but no definite information on possible condensation temperatures exists for this nucleation path. A study of the details of such models is beyond  the scope of this paper.

\subsubsection{Iron condensation temperature}

The appearance temperature of iron dust given in Table~\ref{TabDustModl} may appear to be rather low. It is to be noted, however, that the growth coefficient $\alpha$ for metals like iron is typically of the order of unity, while that measured for silicates is of the order of 0.1 \citep[cf. the discussion in][]{Gai99,Gai13}. This means that the growth of iron is more efficient than that of silicates up to much smaller densities in the outflow because the essential growth species (SiO or Fe) have comparable particle concentrations in the gas phase. This, in turn, means that iron growth extends to significantly lower temperatures than silicate growth. Preliminary results of a more detailed model of iron dust condensation (in preparation) suggest that indeed large quantities of iron dust could be formd at low temperatures.

It has to be observed also that the apparently featureless continuous emission component seen in the spectra of supergiants, which is attributed by us to iron dust, may at least in part also be due to free-free emission of a chromosphere \citep{Gil74,Skin88} or even to the formation of some carbonaceous material \citep[for instance ][]{Syl98,Ver09}. The true physical meaning of the appearance temperature of the continuous opacity source derived from spectral fitting therefore is presently not completely clear. 


\section{Dust condensation}

\label{SectDuGrMod}

In the preceding sections we did not care about how dust forms in the outflow. Now we study how models for the growth of dust under circumstellar conditions can reproduce the results for the dust quantities derived from models of the infrared emission. 

\subsection{Model assumptions}

We restrict our consideration to the three stars $\mu$~Cep, RW Cyg, and RS Per. For these stars the infrared emission could be modelled with good accuracy by a spherically symmetric stationary outflow driven by radiation pressure on dust, resulting in a reasonable set of dust parameters, and at the same time the model results suggest that the silicate dust forms to a considerable fraction by precipitation on top of alumina grains, at least in case of the stars considered by us. This case is amenable to modelling the dust formation in the outflow from a massive supergiant by a simple model for a dust driven wind.

It is assumed that the envelope of the star consists (i) of an interior zone where convective cells and turbulence dominate the motion of matter, but with low systematic outflow velocity, and (ii) of an outer envelope, where silicate condensation and ensuing radiation pressure on dust trigger a systematic outflow with highly supersonic velocity. In the inner part of the envelope alumina dust forms close to the star and is moved around by convection and turbulence. If an outward moving parcel of the gas enters the distance region where the alumina grains are sufficiently cool that silicate starts to precipitate at their surface, the gas parcel is driven out by radiation pressure. Due to the random character of the flow in the inner part of the envelope the dust condensation and acceleration of matter will also occur in a random fashion in space and time. The border between the two regions of the envelope will be highly fuzzy because of a side by side existence of regions with varying density and dust composition. For simplicity we average this complex flow and most likely highly clumpy density structure and assume that the resulting matter and momentum transport equations equal that of a stationary, spherically symmetric outflow.
  
The formation of alumina dust in the inner part of the envelope close to the star is not considered. It is assumed that this dust is dragged by the turbulent flow into the border region between the inner and the outer envelope where silicates start to precipitate on the alumina grains which serve as seed particles. 

With respect to the growth process of silicates it is assumed that it is dominated by  species from the gas phase that are not ionised. This is no matter of course since massive supergiants of early spectral type M possess chromospheres that are sources of ionising UV radiation \citep{Skin88}. For example, the investigation of the chemistry in the envelope of $\alpha$ Ori by \citet{Bec92} demonstrated that in particular the atoms of the dust forming elements Si, Mg, and Fe are fully ionised by the UV output from a warm chromosphere in that case. It is also for this reason that we do not consider the objects $\alpha$~Ori and $\alpha$~Sco (a binary with a B star companion). Observationally, an anticorrelation between the 10 $\mu$m excess do to dust and the Ca H and K emission lines from a chromosphere is found in the sense that no dust seems to be formed in stars with a chromosphere and dust is found in stars without indication for the presence of a chromosphere, with $\alpha$ Ori with little dust formed being a transition case \citep{Jen72,Hag83,Gla87}. 

Our test objects seem not to show indications for chromospheric emission and we assume therefore that the molecular species responsible for the silicate growth are neutral. The dominating species bearing the dust-forming elements Si, O, Mg, and Fe under the pressure and temperature conditions prevailing in a circumstellar shell then would be SiO, H$_2$O, Mg, and Fe. The net chemical reaction for the formation of an iron-bearing silicate with olivine composition from these species is
\begin{equation}
2x\,\mathrm{Mg}+2(1-x)\,\mathrm{Fe}+\mathrm{SiO}+3\mathrm{H}_2\mathrm{O} \to \mathrm{Mg}_{2x}\mathrm{Fe}_{2-2x}^{\,}\mathrm{SiO}_4+3\mathrm{H}_2\mathrm{O}\,.
\label{GrowthStep}
\end{equation}
We calculate the growth process as described in \citet{Fer06}, but assume a fixed value of $x=.5$ for simplicity. 

In the recent past it has been claimed that molecules with composition MgSiO$_3$ and Mg$_2$SiO$_4$ might also be abundant and contribute to dust formation \citep{Mau18,Val20}, but such molecules are not found in the gas phase in vaporisation experiments for silicates \citep[for instance ][]{Nic95,Cos17} and it is not yet  demonstrated that they really exist. We do not include such molecular compounds in our calculation.   

\subsection{Equations}

The calculation follow the concepts outlined in \citet{Fer06} with some modifications. More details can be found in \cite{Gai13}.

\subsubsection{Gas velocity}

The flow of the dusty gas is treated in the tight coupling approximation which assumes that the stopping time of dust particles due to collisions with particles from the carrier gas is short enough that a steady state drift velocity of dust grains relative to the gas, $w_\mathrm{dr}$, is obtained. This enables a separate calculation of the gas velocity and the drift velocity. Further it is assumed that the dust shell is optically thin such that the radiation pressure is due to the unshielded stellar radiation, and only one dust component, the silicate, is responsible for the radiative acceleration of the dusty gas. The velocity of the gas component, $v$, is determined by Eq.~(\ref{EqVel}), which reads in the present case as follows:
\begin{equation}
v\,{\mathrm{d}\,v\over\mathrm{d}\,r}= -{GM_*\over r^2}+{L_*\over 4\pi c r^2} \ \varkappa_{\mathrm{pr}}\cdot f\,.
\label{EqVel1} 
\end{equation}
Here $M_*$ and $L_*$ are the stellar mass and luminosity, $c$ is the velocity of light, and $\varkappa_{\mathrm{pr}}=\kappa^\mathrm{pr}_{1,sil}\psi_\mathrm{sil}$ is the flux-averaged dust opacity of the silicate dust per unit mass of the dusty gas, defined by Eq.~(\ref{DefKapPrMix}). The quantity $f$ is the fraction of the key element for silicate dust condensation (chosen to be Si), which is determined by an equation for the growth of dust grains.

Equation~(\ref{EqVel1}) can only be applied to that part of the stellar outflow where the outflow velocity is supersonic, otherwise we would have to include the pressure force. The cause of the supersonic outflow from the star is thought to be the radiative acceleration on the silicate dust which commences once the temperature in the outflowing gas drops below some critical temperature. Acceleration of the dusty gas by radiation pressure occurs once the right hand side of Eq.~(\ref{EqVel1}) becomes $>0$ which requires in our case the degree of condensation, $f$, to exceed the value
\begin{equation}
f_\mathrm{min}={4\pi cGM_*\over L_*}\,.
\end{equation}      
The initial growth of the dust up to $f=f_\mathrm{min}$ occurs at subsonic or nearly sonic velocities. For simplicity we assume for this part of the flow 
\begin{equation}
{\mathrm{d}\,v\over\mathrm{d}\,r}=0\quad\mathrm{for}\ f<f_\mathrm{min}\,,
\end{equation}
that is we assume a constant velocity equal to some initial value $v_0$ as long as $f<f_\mathrm{min}$, and determine $v$ by integrating Eq.~(\ref{EqVel1}) in the range $f>f_\mathrm{min}$.
   
\begin{table*}

\caption{Experimental determinations of sticking coefficient $\alpha$.}

\begin{tabular}{lllll}
\hline
\hline
\noalign{\smallskip}
Author & \multicolumn{1}{c}{$\alpha$} &  Atmosphere & \multicolumn{1}{c}{$T$} & Process \\
\noalign{\smallskip}
\hline
\noalign{\smallskip}
\citet{Tac14}   & 0.025 \dots\ 0.12 & H$_2$, H$_2$O & 1160         & condensation \\
\citet{Tac15}   & $0.038\pm 0.005$  & H$_2$, H$_2$O & 1235         & condensation \\
\citet{Has90}   & 0.05 \dots\ 0.1   & vacuum        & 1800 -- 2100 & evaporation  \\
\citet{Nag96}   & 0.06 \dots\ 0.2   & H$_2$         & 2000         & evaporation  \\
\citet{Tsu98}   & 0.04 \dots\ 0.2   & H$_2$         & 1500 -- 1700 & evaporation  \\
\noalign{\smallskip}
\hline
\end{tabular}

\label{TabCondCoeff}
\end{table*}

\subsubsection{Drift velocity}

We include the effects of particle drift on grain growth in our model because particle drift relative to the gas induced by different efficiencies of radiation pressure on gas and dust becomes very important for red supergiants at all mass-loss rates. The details of our approach to calculate the drift velocity $w_\mathrm{dr}$ are described in \citet{Gai13}. 

In a stationary state the drift velocity is such that the driving force for the
motion of a dust grain through the gas is just cancelled by the frictional force (the tight coupling approximation), which means
\begin{equation}
w_\mathrm{dr}=\tau_{\rm s}{GM_*\over r^2}\left(\Gamma_\mathrm{d}-1\right)\,.
\label{EqDefFric}
\end{equation}
The quantity $\tau_\mathrm{s}$ is the stopping time of grains by the frictional force due to gas-dust particle collisions,
\begin{equation}
\Gamma_\mathrm{d}={L_*\over 4\pi cM_*}\,{\bar{\sigma}_\mathrm{d}\over m_\mathrm{d}}\,,
\end{equation}
the ratio of radiative acceleration to gravitational pull, $\bar{\sigma}_\mathrm{d}$ the flux-averaged pressure cross section, and $m_\mathrm{d}$ the mass of a dust grain. For spherical grains we have
\begin{equation}
{\sigma_\mathrm{d}\over m_\mathrm{d}}={3\over 4\rho_\mathrm{sol}}\,{\bar{Q}^\mathrm{pr}\over a}\,,
\end{equation}
where $\rho_\mathrm{sol}$ is the bulk density of the condensed phase. The momentum gain of the grains from the radiation field is transferred by gas-dust collisions to the gas and the total radiation pressure on all grains in a dusty gas appears as driving force on the gas-dust mixture (Sect.~\ref{SectWiMo}).

The stopping time $\tau_\mathrm{s}$ depends in an intricate way on the drift velocity \citep[for instance ][]{Gai13}. \citet{Kwo75} has proposed to calculate this by a procedure that interpolates between the limit forms of the stopping time for subsonic and supersonic drift. In that approximation
\begin{equation}
\tau_{\rm s}=\tau_{\rm s,0}\left(1+{9\pi\over64}S^2\right)^{-{1\over2}}\,,
\label{TauFricIpo}
\end{equation}
where
\begin{align}
\tau_{\rm s,0} &={\sqrt\pi\over2}\,{a\over v_{\rm s}}\,
{\rho_\mathrm{sol}\over\varrho_{\rm g}} \label{DefCollTau0}
\end{align}
is the stopping times for subsonic drift and
\begin{align}
S&={w_\mathrm{dr}/v_{\rm s}}\\
v_{\rm s}&=\left(2kT_{\rm g}/m_{\rm g}\right)^{1/2}\,,
\end{align}
with $a$ being the particle radius, and $m_\mathrm{g}$ the average mass of a gas particle \citep[see, for example, ][ for details]{Gai13}. This interpolation formula approximates the exact result for the $w_\mathrm{dr}$-dependence of $\tau_{\rm s}$ within a few percent accuracy.

The drift velocities depend on the particle radius $a$. As $\tau_{\rm s,0}\to0$
for $a\to0$, very small particles are strongly coupled to the gas and do not
move relative to the gas. This holds in particular during the phase of
incipient particle growth where particle sizes are only of the order of 
nano-metres. If they grow bigger they start to move through the gas and may
acquire strongly super-sonic drift velocities. This has to be considered if dust
growth is calculated because the relative motion with respect to the gas 
increases the frequency of collisions with growth species. This enhancement of growth efficiency is negligible or at most moderately important for AGB stars, but becomes significant for massive supergiants with their high radiative flux.

\subsubsection{Temperature}

For the radial variation of the gas temperature we assume for simplicity a grey temperature distribution
\begin{equation}
T^4(r)={1\over2}T_\mathrm{eff}^4{R_*^2\over r^2}\,.
\end{equation}
We also neglect for our present calculations all possible differences between gas temperature and lattice temperature of dust (though this is accounted for in our models for the dust spectrum). 

\subsubsection{Grain growth}

We aim to determine the growth of silicate dust on pre-formed seed nuclei which we assume to be spherical particles of alumina  with radius $a_0$. The radius with which the alumina grains enter the zone where silicate material starts to precipitate on top of the alumina grains is not known, nor do we know the concentration, $\epsilon_\mathrm{gr}$, of such seed particles. For simplicity we assume that all seed particles have the same radius $a_0$. For the concentration  $\epsilon_\mathrm{gr}$ see later.

The change of radius $a$ of a spherical dust grain is
\begin{equation}
{\mathrm{d}\,a\over\mathrm{d}\,t}=V_1\left(J^\mathrm{g}-J^\mathrm{d}\right)\,,
\label{EqGrGr}
\end{equation}
where $V_1$ is the contribution to the volume of the grain by a single nominal molecule from which the dust material is built, $J^\mathrm{g}$ is the collision rate per unit area of grain surface with the rate determining molecule from the gas phase that is successfully integrated into the growing grain, and $J^\mathrm{d}$ is the rate of nominal molecules getting lost from the grain by destructive processes per unit area of grain surface. 

The volume $V_1$ is given by
\begin{equation}
V_1={A_\mathrm{sol}\,m_\mathrm{H}\over \nu_\mathrm{key}\rho_\mathrm{sol}}\,,
\label{EqVolFormUn}
\end{equation}
where $A_\mathrm{sol}$ is the atomic weight of the nominal molecule from which the dust material is built, $\rho_\mathrm{sol}$ is the mass density of the dust material, $\nu_\mathrm{key}$ is the number of key elements in the formula unit, and $m_\mathrm{H}$ the atomic mass unit. The corresponding data are given in Table~\ref{TabCoSu}.

The quantity $J^\mathrm{g}$, called the growth current density, is given by
\begin{equation}
J^\mathrm{g}=J^\mathrm{g}_\mathrm{nd}\Phi\left(w_\mathrm{dr},T_\mathrm{g}\right)\,,
\end{equation}
where $J^\mathrm{g}_\mathrm{nd}$ is the growth current density for a dust grain at rest with respect to the gas phase (no drift), and $\Phi\left(w_\mathrm{dr},T_\mathrm{g}\right)$ is a correction factor that accounts for a drift of the grain with velocity $w_\mathrm{dr}$ with respect to the gas. 

The growth current $J^\mathrm{g}_\mathrm{nd}$ for a grain co-moving with the gas is
\begin{equation}
J^\mathrm{g}_\mathrm{nd}=\alpha\,n_\mathrm{g}\sqrt{kT_\mathrm{g}\over2\pi A_\mathrm{g}m_\mathrm{H}}\,.
\end{equation}
Here $\alpha$ is the probability that a SiO molecule is attached to the surface and builds a new formula unit of the solid instead of being desorbed again or merely being scattered in a molecule-grain collision event, $n_\mathrm{g}$ is the gas phase particle density of SiO molecules, $A_\mathrm{g}$ is the atomic weight of the key molecule from the gas phase for particle growth (SiO) and $T_\mathrm{g}$ the gas temperature. 

The sticking coefficient, $\alpha$, for silicate growth was the subject of a number of experimental investigations which are summarised in Table \ref{TabCondCoeff}. Its value depends on the particular kind of experiment performed and varies considerably for different experimental runs. The temperature at which the experiments where run was in all cases much higher than the temperature for which silicate condensation occurs under circumstellar conditions. Though the available experimental data do not show a clear temperature dependence of $\alpha$, this may be different at lower temperatures because the growth of silicates requires surface diffusion and reactions of the molecular components and intermediate products that finally form the building block of the silicate. Typically the value of $\alpha$ is of the order of 0.1 with an accuracy within a factor of about two and no clear temperature dependency. This is the value which was used in our previous work on dust condensation \citep[for instance ][]{Gai99,Fer06}. Here we consider $\alpha$ as a free parameter which is varied to fit the results of a condensation calculation with observational properties of the dust shell. This serves to check if the value of $\alpha$ obtained in this way is compatible with the results for $\alpha$ from laboratory experiments.

The correction factor for particle drift, $\Phi$, is calculated for spherical grains and for the case of specular reflection of H$_2$, He at the surface of a dust grain. Details for this are given in \citet{Gai13}. A convenient interpolation formula for $\Phi\left(w_\mathrm{dr},T_\mathrm{g}\right)$ between the limit cases of no drift at one hand and of super-sonic drift with respect to the gas at the other hand is
\begin{equation}
\Phi\left(w_\mathrm{dr},T_\mathrm{g}\right)=\sqrt{1+{\pi A_\mathrm{g}m_\mathrm{H}\over 8kT_\mathrm{g}}w_\mathrm{dr}^2}\,,
\end{equation}
which is accurate within a few percent in the transition region.

The grain growth is counteracted by a number of processes, for instance by thermal vaporisation, chemical reactions with ambient gas, sputtering by impinging gas-phase species at high-velocity grain drift, sputtering by hot ambient plasma, photo desorption by high energy photons. In the relatively cool environment in the dust-forming zone of a supergiant the two most important processes probably are thermal vaporisation and sputtering induced by grain drift. The destruction current $J^\mathrm{d}_\mathrm{nd}$ by thermal decomposition is
\begin{equation}
J^\mathrm{d}=\alpha\,{p^\mathrm{eq}_\mathrm{g}\over \sqrt{2\pi A_\mathrm{g}m_\mathrm{H}\,kT_\mathrm{d}}}\,.
\label{RatDustVap}
\end{equation}
Here $T_\mathrm{d}$ is the dust temperature and $p^\mathrm{eq}_\mathrm{g}$ is the partial pressure of the key molecule at chemical equilibrium between the gas phase and the solid. A destruction by drift-induced sputtering is neglected in our calculation because drift velocities turn out to be less than about the threshold for efficient sputtering of about 50 km\,s$^{-1}$ \citep[cf.][]{Kru97}. For the calculation of $p^\mathrm{eq}_\mathrm{g}$ see \citet{Gai99} and \citet{Fer02}. What we would need for this calculation are data for the thermodynamic properties of the amorphous materials, which seem not to be available. Therefore we used data for the corresponding crystalline materials, which somewhat overestimate the stability of the material against chemisputtering.   

The silicate dust is only stable against vaporisation if conditions are such that the right hand side of Eq.~(\ref{EqGrGr}) is $>0$, which means: at sufficiently low temperature. As long as the right hand side is $<0$ we held the particle radius $a$ fixed at the initial value $a_0$ of the seed particle. Once the right hand side becomes $<0$, we determine $a$ by integrating the differential equation.   

\subsubsection{Particle density}

The progress of condensation is characterised by the fraction $f$ of the key element of a dust species condensed into dust component. This is can be calculated from
\begin{equation}
f={4\pi\left(a^3_{}-a_0^3\right)\over 3V_1}\,{\epsilon_\mathrm{gr}\over\epsilon_\mathrm{el}}
\,.
\label{DefCondFra}
\end{equation}
because in our model all dust particles at a certain radial distance $r$ have the same particle radius $a$ if they all start from the same initial radius $a_0$. In Eq.~(\ref{DefCondFra}) $\epsilon_\mathrm{el}$ is the elemental abundance relative to H of the key element in the stellar wind (Si for silicate dust), and $\epsilon_\mathrm{gr}$ the number of seed particles per H atom in the outflow onto which the dust material precipitates. 

With this, the particle density $n_\mathrm{g}$ of the growth species is
\begin{equation}
n_\mathrm{g}=\epsilon_\mathrm{el}\left(1-f\right)N_\mathrm{H}\,,
\end{equation}
where
\begin{align}
N_\mathrm{H} &= {\varrho\over\left(1+4\epsilon_\mathrm{He}\right)m_\mathrm{H}}\\
\varrho &= {\dot M\over4\pi r^2v}\,.
\end{align}
The quantity $\dot M$ is the mass-loss rate of the stellar wind.

If alumina particles serve as seed particles then we have in analogy to Eq.~(\ref{DefCondFra})
\begin{equation}
\epsilon_\mathrm{gr}=f_\mathrm{Al}\epsilon_\mathrm{Al}\ V_{1,\mathrm{al}}/(4\pi a_0^3)\,,
\label{DefConcSeed}
\end{equation}
where $f_\mathrm{Al}$ is the fraction of Al condensed into alumina grains and $V_{1,\mathrm{al}}$ is given by Eq.~(\ref{EqVolFormUn}) for the case of alumina as condensed phase.

\subsection{Model calculation}

We calculate models of the silicate dust condensation for the stars $\mu$ Cep, RW Cyg, and RS Per. The stellar parameters $M_*$, $L_*$, and $T_\mathrm{eff}$ are set to the values derived from observations (given in Table \ref{TabStarParm}). The mass-loss rate $\dot M$ is set to the value found from the fit between observed and model spectra for the accelerated outflow model given in Table \ref{TabCompFit}. This value and not the value derived from CO or atomic line widths is used to achieve consistency with the optical properties of the dust absorption model. For the opacity $\varkappa_\mathrm{pr}$ we calculate the Planck-average with effective temperature of the star using optical data of our amorphous non-stoichiometric silicate material with composition $x=0.5$ and typical element abundances for supergiants from Table~\ref{TabElAbu}.

The coupled set of equations (\ref{EqVel1}) and (\ref{EqGrGr}) for the outflow velocity of the gas and for the grain growth, respectively, augmented by the equations for calculating the drift velocity, $w_\mathrm{d}$, and the degree of condensation, $f$, are solved numerically, starting at some radial distance $r_0$ with an initial velocity $v_0$ and an initial grain radius $a_0$. The radial distance $r_0$ is chosen close enough to the star such that at the corresponding temperature no silicate dust exists. The initial velocity $v_0$ for the outflow is chosen as 2 km\,s$^{-1}$ as in the optimum fit models. This is close to the sound velocity at the inner edge of the dust shell. The point of onset of silicate dust condensation is found during the course of the calculation when the right hand side of Eq.~(\ref{EqGrGr}) becomes $>0$ for the first time.

It is assumed that the alumina particles formed in the region between the stellar atmosphere and the inner edge of the silicate dust shell act as seed particles for the growth of silicate dust grains. The fraction of Al condensed into such grains is set to the value found by fitting the observed spectrum with model spectra (Table \ref{TabModFcond1}). The concentration $\epsilon_\mathrm{gr}$ of the seed grains is given by Eq.~(\ref{DefConcSeed}) if we specify the radius $a_0$. This radius of alumina grains is not really fixed by the radiative transfer models. Though this enters as an important parameter into the opacity calculation, the quality of the fit of the mid-infrared spectrum only weakly depends on the assumed size of the alumina dust grains such that no unequivocally preferred typical particle size can be determined. This is because all the grains are small compared to the wavelengths of mid-infrared radiation which is used to compare models with observations, and in the small-particle limit the extinction properties of particles do not depend on the particle size. Particle sizes of grains formed close to the photosphere of giants have been derived from observations of scattered radiation by \citet{Sci15} and \citet{Hau19} to be as large as 0.3\ \dots\ 0.5 $\mu$m. These observations refer, however, to a region in the stellar envelope short above the stellar atmosphere. At the inner edge of the silicate dust shell at much larger distances the size spectrum of grains may be different because of settling and mixing in the stellar envelope where convective flows and turbulence prevail. Additionally, no model calculations for the alumina dust condensation in the envelope of supergiants exist, that may give some hints onto expected grain sizes. The radius $a_0$ therefore is considered here as a free parameter. 

\begin{table}

\caption{Dust growth model}

\begin{tabular}{lcccl}
\hline
\hline
\noalign{\smallskip}
Quantity & $\mu$ Cep & RW Cyg & RS Per & Units \\
\noalign{\smallskip}
\hline
\noalign{\smallskip}
$\varkappa_\mathrm{pr}$ & 18.0  & 17.1  & 16.7  & cm$^2$g$^{-1}$ \\
$f_\mathrm{min}$        & 0.034 & 0.079 & 0.084 & \\[.2cm]
$\alpha$                & 0.16  & 0.15  & 0.20  &    \\
$a_0$                   & 0.02  & 0.02  & 0.02  & $\mu$m \\[.2cm]
$T_\mathrm{c}$          & 1091  & 1105  & 1093  & K \\
$T_\mathrm{acc}$        &  938  & 940   & 911   & K \\
$T_\mathrm{app}$        &  910  & 938   & 923   & K \\[.2cm]
$f_\mathrm{Si}$ (growth) & 0.068 & 0.140 & 0.142 & \\
$f_\mathrm{Si}$ (optim.) & 0.084 & 0.147 & 0.140 & \\
$a_\infty$              & 0.043 & 0.037 & 0.036 & $\mu$m \\
$w_\infty$              & 16.9  & 11.6  & 16.4  & km\,s$^{-1}$ \\
$v_\infty$              & 20.0  & 23.0  & 20.0  & km\,s$^{-1}$ \\
\noalign{\smallskip}
\hline
\end{tabular}

\label{TabCondModel}
\end{table}

The sticking coefficient $\alpha$ for silicate grain growth is also considered as a free parameter, as already mentioned above. The growth model therefore depends on two parameters. It is required that the calculated outflow velocity at a distance of $10^4$ $R_*$ equals the observed wind velocity, $v_\infty$. The two parameters $\alpha$ and $a_0$ are varied to obtain a reasonable fit for all three comparison stars $\mu$ Cep, RW Cyg, and RS Per with a single value for the seed particle radius $a_0$. This is presently done by hand and inspection of results because we feel that the simple kind of condensation calculation does not justify to use sophisticated minimisation methods at the present stage of modelling. 

Figure \ref{FigModGrowthMuCep} shows for the case of $\mu$ Cep the model results for the radial variation of the degree of condensation of Si into silicate dust, $f$, the gas velocity, $v$, and the drift velocity of the silicate dust grains, $w_\mathrm{d}$. The structure of the gas velocity of our wind model which accounts for the details of dust condensation and wind acceleration is already well represented by the simple analytic wind model defined by Eq.~(\ref{EqFlowSimpl}). Furthermore, the variation of the degree of condensation is not far from the step-like variation of the instantaneous condensation assumption on which the simplified model is based. This lends some credit to the reliability of the properties of the dust shells derived in Sect.~\ref{SectModl} on the basis of the simplified wind model. Of particular interest are the final degree of condensation of Si into silicate dust and the  temperature for onset of massive dust condensation. Only if we achieve consistency of the values obtained from an analysis of the infrared spectrum and the results of the growth model it can be supposed that it is possible to construct quantitatively correct dust formation models. The results of our preliminary fit for all three stars are shown in Table~\ref{TabCondModel}. 

It is found that a choice of $a_0=0.02\ \mu$ allows a reasonably good fitting of the models for all three stars. This radius is much smaller than the assumed radius of 0.2 $\mu$m in the opacity calculations. This means that we assume that the alumina particles show a broad distribution of sizes where most of the mass is contained in a small fraction of big grains from the upper end of the size distribution, while the majority of alumina grains are much smaller and serve as the dominating population of seed particles for the silicate grains. Our assumption then is that of a size distribution decreasing with increasing size, but not so steep that also most of the dust mass is also contained in the smallest particles (as, for instance in the MRN size distribution widely assumed in model calculations of circumstellar dust shells). Whether this assumption is realistic could only be checked by a consistent model of simultaneous formation and growth of alumina and silicate dust which is presently not available and out of the scope of the present work.

The corresponding sticking coefficient $\alpha$ required for a match between calculated and observed outflow velocity of the gas, $v_\infty$, is well within the range of values found in laboratory experiments, though at the upper end of the range of the experimental data. This means that it is possible to achieve full consistency between dust and gas mass-loss rates derived from analysis of infrared spectra and a model calculation of the condensation for the main dust component responsible for the acceleration of the dusty gas to the observed highly supersonic outflow velocity, if we assume that the rate determining reaction step for growth of macroscopic silicate dust grains is the reaction indicated in Eq.~(\ref{GrowthStep}) and using data for the growth coefficient $\alpha$ as determined by laboratory experiments for this kind of reaction.  

At the same time, the final degree of condensation, $f_\mathrm{Si}$, at large distance resulting for the condensation model equals the corresponding value obtained by spectral fitting within 20\% in the case of $\mu$ Cep and within less than 5\% in the other two cases. This additionally demonstrates, that a growth model based on the assumption that reaction (\ref{GrowthStep}) is the rate determining growth step and using sticking coefficients as determined for this reaction by laboratory experiments, allows for reliable modelling of dust condensation. 
   
Table~\ref{TabCondModel} also shows the temperature $T_\mathrm{acc}$ where the degree of condensation $f$ equals the value $f_\mathrm{min}$ for which radiation pressure on the silicate dust exceeds gravitational pull by the star. From that point on the flow is strongly accelerated to supersonic velocities. For all three stars this temperature is close to the appearance temperature, $T_\mathrm{app}$ (also shown in Table~\ref{TabCondModel}), derived from the modelling of the infrared emission. Hence, also with respect to temperature at the inner edge of the silicate dust shell we obtain reasonable consistency between our wind model and the results derived from analysis of infrared emission. 

The temperature $T_\mathrm{c}$ in Table~\ref{TabCondModel} is that temperature at which the rate of precipitation of silicate material on the surface of seed particles exceeds its evaporation rate for the first time in the models. This temperature is ca. 150 K higher than the temperature $T_\mathrm{acc}$. It equals the stability limit of crystalline olivine in an environment with cosmic element abundances because we used thermodynamic data for this material to calculate the chemisputtering rate $J^\mathrm{d}$ (which is the reverse reaction of reaction (\ref{GrowthStep})). Though the final dust material is dominated by amorphous material, it may well be that the initial growth at high temperature results in a crystalline material which at lower temperatures is overgrown by amorphous material, as is discussed in \citet{Gai99}. A more accurate calculation of the early stages of the growth process requires a more realistic calculation of the velocity in the range of subsonic to sonic average outflow velocities, which is out of the scope of this paper.
 
\begin{figure*}

\includegraphics[width=.33\hsize]{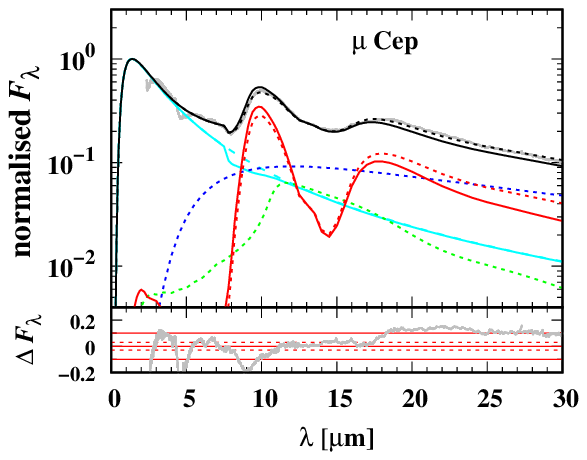}
\includegraphics[width=.33\hsize]{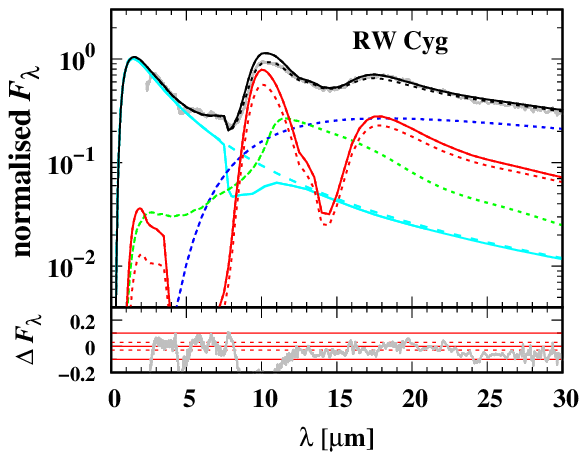}
\includegraphics[width=.33\hsize]{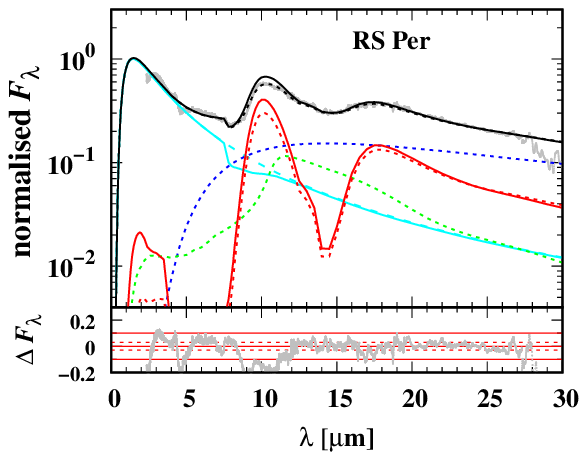}

\caption{
Comparison of observed spectra with synthetic spectra calculated for the optimised accelerated wind model (dashed lines) and for the model considering silicate dust growt (solid lines). Black lines: total spectrum. Red lines: contribution of the silicate component. Grey dots: observed spectrum.
}

\label{FigComModls}
\end{figure*}

The final grain size at large distance, $a_\infty$, following from the growth model is relatively small, see Table~\ref{TabCondModel}. Little is known from observations about grain sizes in outflows from massive supergiants. The few published studies \citep{Pap83,Sno87, Jur96,Sci15} seem to indicate the presence of grains with sizes of the order of $a\sim0.1\ \dots\ 0.3\ \mu$m which is somewhat bigger than what follows from our calculation. Such a bigger size, however, may easily result from agglomeration of grains due to their mutual relative motions by their drift through the carrier gas and/or from the clumpy structure of the outflow resulting in variable growth conditions. A small extent of agglomeration is also suggested by the requirement of non-spherical grains to model the shape of the two strong silicate features of circumstellar dust while production of nano-particles in flames show that amorphous grains formed by precipitation from the gas phase are near-perfect spheres; the symmetry is lost only by agglomeration (see also Sect.~\ref{SectSizeShape}). Both effects leading to bigger grains are neglected in the present simple growth model, but such effects need to be included in future calculations. 

\begin{table}

\caption{Modified dust growth model}

\begin{tabular}{lcccl}
\hline
\hline
\noalign{\smallskip}
Quantity & $\mu$ Cep & RW Cyg & RS Per & Units \\
\noalign{\smallskip}
\hline
\noalign{\smallskip}
$\alpha$                & 0.20  & 0.20  & 0.20  &    \\
$a_0$                   & 0.02  & 0.01  & 0.01  & $\mu$m \\[.2cm]
$T_\mathrm{c}$          & 1085  & 1084  & 1085  & K \\
$T_\mathrm{acc}$        &  886  & 1026  & 1028  & K \\
$f_\mathrm{Si}$ (growth) & 0.071 & 0.24 & 0.22  & \\
$a_\infty$              & 0.039 & 0.020 & 0.015 & $\mu$m \\
$w_\infty$              & 18.9  & 12.4  & 13.2  & km\,s$^{-1}$ \\
$v_\infty$              & 17.8  & 27.9  & 23.1  & km\,s$^{-1}$ \\
$f_\mathrm{gr}$         &  2.0  & 1.0  & 5.0 & \\
\noalign{\smallskip}
\hline
\end{tabular}

\label{TabGrowtModel}
\end{table}

\subsection{Synthetic spectra}

The growth model just considered cannot immediately be integrated in a consistent way into our model program for fitting synthetic and observed spectra since we have presently no growth models for the alumina and iron dust components. In order to get at least a qualitative impression how the results of a growth model for the silicate component fit into the frame of our optimised dust shell models we take the results for the run of dust density and dust temperature of the alumina and iron dust components from an optimised model of our three comparison stars and add the silicate component from a growth model using the mass-loss rate of the gas component. Since we assume that the silicate dust grows on alumina grains as seed particles, we have to calculate the decomposition rate, defined by Eq.~(\ref{RatDustVap}), at the onset of silicate precipitation with the temperature of the alumina grains because of the close thermal coupling of the silicate material at the initial formation of a silicate layer on an aluminium substrate. As significant amounts of silicate overgrows the corundum cores, the temperature of the compound grains shifts to close the temperature of pure silicate grains (since the silicate mantle then dominates the dust absorption, see Fig. \ref{FigAluPliCM}), but the details of this are not considered and for calculating the emission from the grains we simply assume that their temperature equals that of pure silicate grains. The sticking coefficient $\alpha$ is taken to be 0.2 as suggested by the exploratory calculation before, and for the radius $a_0$ of the seeds we take the value assumed above.

What remains unconstrained by observations is the number of seed particles per H atom. In the preceding model it is assumed that all silicate grains are formed with an alumina core. If we compare a dust growth model with the infrared emission from a real dust shell we must admit that also other sources of seed particles could be important, for instance a contribution of direct nucleation of silicates from the gas phase. We therefore take the number of seed particles as a free parameter and vary this to obtain a close fit to the final gas outflow velocity. Technically we realize this by letting the factor $f_\mathrm{Al}$ in Eq.~(\ref{DefConcSeed}) be a free parameter which now may even exceed unity if there are significant other sources for seed particles. It turned out that, at least for the type of model considered here, no good fit seems to be possible if only alumina particles serve as seeds. What we require are numerous small seeds. The results for the best model we could presently obtain are shown in Table~\ref{TabGrowtModel}. The results seem to indicate that in the three objects $\mu$ Cep, RW Cyg, and RS Per also direct nucleation silicate from the gas phase is at work. The final grain size obtained here seems to be rather small compared to what was obtained in the few published estimates for grain sizes of supergiants \citep[$\sim0.1\ \mu$m, cf. ][]{Pap83,Sno87, Jur96,Sci15}, but this may change once agglomeration is considered in the dust growth model.  

Figure \ref{FigComModls} compares the results for the synthetic spectrum of this kind of model with the synthetic spectrum of our previus optimised models and the observed spectrum. The dashed lines are for the semi-analytic accelerated wind model assuming instantaneous condensation, the full lines are for the model where the gas velocity and the degree of condensation for silicate dust are determined by solving the corresponding set of equations for this problem. As can be seen, a reasonable agreement between the two kinds of models can be achieved, though not a perfect one. A perfect agreement cannot be expected, since the real growth process depends on a number of details like the precise variation of gas and dust temperature, evolution of particle size distribution, nature of the seed particle formation process, variation of the iron content of the silicate material with growth conditions, and so forth, which are only cursory treated or neglected in the present model. Such detailed knowledge is not required for the semi-analytic model but is condensed in that model into the few free parameters of the model that are derived from spectral fitting. Nevertheless, with the sole knowledge of the rate-determining step of the silicate growth, laboratory determined values of the growth coefficient and of the thermochemical data of the dust material, opacity data from silicate material formed by vapour deposition on a cold substrate, augmented by an information of the final outflow velocity and mass-loss rate allows us to calculate the contribution of the infrared emission of silicate dust to the spectrum which turns out to be close to what one infers from an optimised radiation transfer modelling. To obtain better agreement it would be necessary to develop a consistent modelling of the dust growth for all of the involved major dust components, which is beyond the scope of this paper, but is planned as a next step. 

\subsection{Summary}

What we see from our results is the following. It is possible to achieve a consistent picture for the condensation of silicate dust for two kinds of models:

First, a radiative transfer model based two ingredients: first, on observations of the infrared spectrum from the dust shell, augmented by a knowledge of the outflow velocity (for instance from molecular lines), of the stellar luminosity and effective temperature, and element abundances; second, on laboratory measured optical constants of the dust material, the nature of which is inferred from the observed spectral features and structure of the spectrum.

Second, a theoretical model for the condensation of silicate dust, also based on two ingredients: First, on a knowledge of stellar luminosity, effective temperature, and element abundances. Second, knowledge of the mass-loss rate for the gas phase and of the optical constants of the dust material, augmented with information on the growth mechanism including the growth coefficient determined by laboratory experiments.

If we take for one and the same star the mass-loss rate of the gas phase resulting from the first kind of model, we can reproduce with the second kind of model the essential properties of the first kind of model. This shows, that the model for silicate dust growth that assumes attachment of SiO molecules from the gas phase being the rate-determining step of the growth process and uses the laboratory-determined growth coefficient is able to predict the amount of silicate dust produced in the outflow. 


\section{Concluding remarks}

We constructed models for the circumstellar dust shells for a set of nine well-studied massive supergiants with optically thin dust shells for which (i) mid-infrared observations by the ISO satellite are available from \citet{Slo03b}, which (ii) have reasonably well known stellar parameters, and for which (iii) mass-loss rates and expansion velocities at large distances of the stellar outflow have been determined from atomic and/or molecular lines. 

Two kinds of models are considered, one with constant outflow velocity but with otherwise unspecified dynamical properties, and a second one for dust driven winds where the outflow velocity is determined by a simplified analytic solution of the momentum equation. Both kind of models make use of the instantaneous condensation approximation which assumes that dust suddenly forms if the radiative equilibrium temperature of dust grains drops below some characteristic value $T_i$ for a dust species $i$. This temperature and the quantities of dust contained in the circumstellar shell are determined by fitting the observed mid-infrared spectra with synthetic spectra calculated for the models.

The models consider three dust components: silicate dust, which is responsible for the $\sim10\ \mu$m and $\sim18\ \mu$m features seen in the spectra, alumina dust, which seems to be required to fill-up the absorption trough between the two silicate features, and iron dust (either as a separate component or as inclusions in silicate grains), which is assumed to be responsible for the continuum excess emission shortwards of $\sim9\ \mu$m wavelength. A possible contribution of a warm chromosphere is neglected. For calculating the silicate dust extinction we use the set of optical constants derived in our previous paper \citep{Tam17}. These data refer to a highly amorphous, non-stochiometric iron-magnesium silicate that has been prepared by vapour deposition on a cold substrate. Its formation condition likely resembles the condition under which silicate dust condenses under circumstellar conditions, such that one can expect that this kind of material bears some similarity to the circumstellar silicate material. 

It is possible to fit the dust emission spectra of our set of stars rather well. In particular for the accelerated-outflow model, the optimum fit models for the dust spectrum fairly reproduce (except for two cases: $\alpha$ Ori and $\alpha$ Sco) at the same time also the gas outflow velocity at large distances and the mass-loss rate obtained in a completely different way by analysis of atomic or molecular lines. Since neither the knowledge of the terminal velocity nor of the mass-loss rate is used in the construction of the dust shell model, the success of the accelerated outflow model in conjunction with the opacity for the amorphous, non-stoichiometric silicate dust material from \citet{Tam17} suggests that indeed the extinction properties of this kind of material are close to that of the real silicate dust in circumstellar shells. 

For most of the studied objects we obtain a good fit only for a rather low initial velocity at the inner border of the silicate dust shell, of the order of the sonic velocity or slightly above. This indicates that for these objects we presently observe a phase of long lasting active dust formation where there only exists a slow average outflow interior of the inner edge of the silicate dust shell and the strong wind acceleration to the highly supersonic terminal wind velocity is related to the formation of the silicate dust component. This kind of model is definitely excluded for PZ Cas. In that case we likely see an old expanding dust shell from an earlier phase with a much higher mass-loss rate than presently, or dust formation in an outflow  driven by a different mechanism, for instance in the case of $\alpha$ Ori. Acceptable, though not optimal, fits can also be obtained with much higher initial velocities for part of the other objects.

For four of the objects ($\mu$ Cep, RW Cyg, W Per, and RS Per) the most satisfactory fit is obtained by the accelerated outflow model with low initial velocity. For three of them ($\mu$ Cep, RW Cyg, RS Per) the dust temperature at the inner edge of the silicate dust shell is found in the model to be somewhat above 900 K, while for W Per it is somewhat above 600 K. In the first three cases the temperature is below, but not much lower than, the upper stability limit for silicate material under circumstellar conditions. We interpret this finding as indication that the silicate dust forms in these stars, at least to a significant fraction, by precipitation on the alumina dust grains which are formed much closer to the star and which can serve as seed particles for the silicate dust growth. For W Per either the simple wind model is not applicable or a different mechanism for formation of seed particles is at work.

We use the model results for the silicate dust around $\mu$ Cep, RW Cyg, and RS Per to check, if we are able to reproduce them by calculating the growth of silicate dust grains in their outflow. The fact that in all likelihood alumina dust particles serve as seed particles for a major fraction of silicate grains in these objects eliminates the necessity to deal with the still not solved problem of nucleation of dust grains and leaves us with the problem of how dust grains collect the condensible material from the gas phase. We integrate the set of equations describing dust  growth under conditions encountered in the envelopes of massive supergiants. The main uncertain parameter for this problem is the growth coefficient $\alpha$, the probability that a collision of a SiO molecule with the surface of a dust grains ends up with its incorporation into the silicate material instead of being merely scattered. This quantity has been studied by a number of researchers in the laboratory under varying conditions with a considerable width of the range of values found for $\alpha$. We therefore turn the tables and ask for which value of $\alpha$ the results of the solution of the coupled set of equations for grain growth and wind driving reproduces the model of the dust shell for each of our three test cases? It turns out, that a value of $\alpha$ from the upper range of values obtained in laboratory experiments fulfils this purpose. With this, the dust mass in the shell, the typical condensation temperature, and the typical radius of circumstellar dust grains (as determined by studies of light scattered by grains in circumstellar envelopes) are largely reproduced. 

From the two major problems of dust formation in stellar outflows,
\begin{enumerate}
\item the formation of seed nuclei at the molecular level, and
\item their growth to macroscopic dust grains,
\end{enumerate}
our results suggest that for the second part of the problem we gradually come close to a quantitative solution of it. If the silicate dust growth in a hydrogen-rich environment is dominated by neutral species, a calculation based on the information that can be inferred from the published results of the appropriate laboratory experiments provides already quantitatively nearly correct answers. 

Clearly, this needs further confirmation by studying more objects. In particular, more objects with sufficiently simple dust shell properties need to be identified and studied in detail, that allow us to study dust formation without the severe complexities of a highly clumpy and variable outflow as it is usually observed and which hampers a disentangling of radiative transport, hydrodynamics, and grain growth effects. The modelling also needs to be extended to include alumina dust formation in the zone interior to the silicate-dominated envelope into the models, which is currently underway.


\begin{acknowledgements}
We express our gratitude to Dr.~Theiss for making available for us the SCOUT software. We are thankful to Dr. S. Wetzel for his technical and scientific supports. We thank Hans-Werner Becker and Detlef Rogalla for Rutherford Backscattering analysis. Our project has been supported by the Deutsche Forschungs\-gemeinschaft (DFG) under the Priority Program SPP1385. This work is (partly) performed at and supported by RUBION, central unit of the Ruhr-Universit\"at Bochum. This research has made use of NASA's Astrophysics Data System.
\end{acknowledgements}


\begin{appendix}

\section{The binary system $\alpha$ Sco}

\label{SecBinAlfSco}

Massive stars from the range of initial masses $11\ \mathrm{M}_{\sun}\lesssim M\lesssim25\ \mathrm{M}_{\sun}$ evolve to the RGB where they become sufficiently cool that they may form dust in their outflows. These stars are usually associated with dust shells only if they are solitary. As members of double stars they mostly form no dust or only small quantities of it \citep{Bus88}, since the strong UV radiation from a nearby hot companion may suppress dust formation. Hence, all stars of our sample, except for $\alpha$ Sco, are single stars. 

For high-mass stars, the fraction of binaries is high, about 50\% to 80\% \citep{Chi13} and solitary red supergiants are the exception. Hence, red supergiants in binary systems with hot companions that do form substantial amounts of dust seem to be rare; in fact, $\alpha$ Sco, also known as Antares, seems to be the sole object of this kind. Such objects are particularly interesting because, if the wind material of the cool component is in front of the hot component, one can directly investigate absorption lines from the cool star's wind in the spectrum of the hot star and derive the depletion of the dust forming elements from the gas phase due to dust condensation \citep{Sno87}.

$\alpha$ Sco is member of a binary system formed by the M1.5 Iab cool supergiant $\alpha $ Sco A and the hot B2.5 V main sequence star $\alpha$ Sco B. The properties of the binary and the mass-loss rate from the cool star were investigated in detail by \citet{Kud78} and \citet{Rei08}. They give masses of 18 and 2.2 $M_{\sun}$ and stellar radii of 575 and $\approx5.2$ R$_{\sun}$ for $\alpha$ Sca A and B, respectively. The distance between binary components is 820 AU. The Roche radius of component A in the binary system calculated from the approximation \citep{Egg83} 
\begin{equation}
R=a\,{0.49\,r^2\over0.6 r^2+\ln(1+r,q)}\,,
\end{equation}
where $r=(M_\mathrm{A}/M_\mathrm{B})^{1/3}$ and $a$ is the binary distance (circular orbit assumed), results to $R=396$ AU. The stellar radius of component A corresponds to 2.7 AU, that is, the supergiant is far from filling its roche lobe. The properties of the dust and of the dusty stellar outflow from component A are determined within a few tens of stellar radii of the M star. They are essentially undisturbed by the gravity field of component B.
 
The UV radiation from the hot star forms a small H II region within the outflow. The properties of the ionised region formed within the outflow of component A were also studied by \citet{Kud78}. The equilibrium position of the ionisation/shock front between the colliding winds from the M and B star on the line connecting the two stars is estimated to be at 320 AU away from the M star. A big fraction of the wind material, thence is occupied by the ionised region and the whole structure is strongly asymmetric \citep{Bra12}. But the inner part of the wind from $\alpha$ Sco A up to a distance of about 100 stellar radii that is essential for the mid-infrared emission from dust remains unaffected by the presence of the ionised region.

Thus, the dusty wind from $\alpha$ Sco A can be treated as if it were an outflow from  a single star. Nevertheless, contrary to the other stars we have for this system some direct observational information of how much material was consumed by the dust formation process. \citet{Sno87} and \citet{Baa07} studied UV absorption lines in the spectrum of the B star due to wind material of the M star in front of the B star and derived the depletion factors of the elements responsible for the formation of the most abundant dust species (O, Mg, Si, Al, Fe). The data presented in \citet{Sno87} indicate a low depletion of the main refractory dust-forming elements Si, Mg, and Fe in the outflow from the M-component, but a significant depletion of Al. The depletion in the outflow is found to be smaller than for the diffuse interstellar medium.  The data presented in \citet{Baa07} do not allow us to draw definite conclusions with respect to the true degree of depletion, but also seem to indicate lower depletions as in the diffuse ISM.   


\section{Optical constants}

\label{AppOptTab}

Numerical values of the optical constants of amorphous silicates in the spectral region between $2\,\mu$m and $200\,\mu$m \citep{Tam17} are shown in Table \ref{TabOptDat4}.

\begin{table*}
\caption{Optical constants data of the four silicate dust materials in the infrared spectra region.}

{\small
\begin{tabular}{rcccccccc}
\hline
\hline
\noalign{\smallskip}
  & \multicolumn{2}{c}{Si-film1} & \multicolumn{2}{c}{Si-film2} & \multicolumn{2}{c}{Si-Film3} & \multicolumn{2}{c}{Si-Film4} \\
\noalign{\smallskip}
  & \multicolumn{2}{c}{Mg\# = 1.00} & \multicolumn{2}{c}{Mg\# = 0.64} & \multicolumn{2}{c}{Mg\# = 0.32} &\multicolumn{2}{c}{Mg\# = 0.00} \\ 
\noalign{\smallskip}
\hline
\noalign{\smallskip}
$\lambda$ &  $n$  &    $k$    &  $n$  &    $k$    &  $n$  &    $k$    &  $n$  &    $k$    \\
\noalign{\smallskip}
\hline
\noalign{\smallskip}
  2.00 & 1.544 &  1.619E-5 & 1.618 &  1.854E-5 & 1.762 &  1.490E-5 & 1.909 &  1.309E-5 \\
  3.00 & 1.527 &  5.790E-5 & 1.597 &  6.951E-5 & 1.745 &  5.503E-5 & 1.895 &  4.486E-5 \\
  4.00 & 1.500 &  1.576E-4 & 1.564 &  1.920E-4 & 1.718 &  1.492E-4 & 1.872 &  1.204E-4 \\
  5.00 & 1.459 &  3.833E-4 & 1.516 &  4.630E-4 & 1.689 &  3.482E-4 & 1.840 &  2.905E-4 \\
  5.50 & 1.431 &  5.933E-4 & 1.483 &  7.156E-4 & 1.670 &  5.198E-4 & 1.817 &  4.484E-4 \\
  6.00 & 1.398 &  9.361E-4 & 1.443 &  1.126E-3 & 1.647 &  7.792E-4 & 1.789 &  7.070E-4 \\
  6.50 & 1.352 &  1.559E-3 & 1.389 &  1.868E-3 & 1.618 &  1.197E-3 & 1.753 &  1.181E-3 \\
  7.00 & 1.291 &  2.894E-3 & 1.317 &  3.497E-3 & 1.580 &  1.941E-3 & 1.705 &  2.249E-3 \\
  7.50 & 1.199 &  6.951E-3 & 1.210 &  8.919E-3 & 1.529 &  3.571E-3 & 1.634 &  5.851E-3 \\
  8.00 & 1.039 &  3.350E-2 & 1.022 &  4.651E-2 & 1.451 &  9.222E-3 & 1.513 &  2.751E-2 \\
  8.20 & 0.947 &  1.055E-1 & 0.916 &  1.052E-1 & 1.406 &  1.678E-2 & 1.441 &  6.055E-2 \\
  8.40 & 0.880 &  2.310E-1 & 0.813 &  2.058E-1 & 1.350 &  3.227E-2 & 1.370 &  1.260E-1 \\
  8.60 & 0.858 &  3.684E-1 & 0.729 &  3.515E-1 & 1.284 &  6.366E-2 & 1.320 &  2.170E-1 \\
  8.80 & 0.857 &  4.954E-1 & 0.697 &  5.257E-1 & 1.212 &  1.171E-1 & 1.305 &  3.207E-1 \\
  9.00 & 0.869 &  6.374E-1 & 0.698 &  7.076E-1 & 1.145 &  1.983E-1 & 1.315 &  4.275E-1 \\
  9.20 & 0.925 &  7.962E-1 & 0.752 &  8.955E-1 & 1.092 &  3.005E-1 & 1.333 &  5.117E-1 \\
  9.40 & 1.036 &  9.397E-1 & 0.859 &  1.072E+0 & 1.059 &  4.199E-1 & 1.359 &  5.906E-1 \\
  9.60 & 1.185 &  1.042E+0 & 1.003 &  1.202E+0 & 1.065 &  5.615E-1 & 1.390 &  6.761E-1 \\
  9.80 & 1.352 &  1.096E+0 & 1.157 &  1.303E+0 & 1.113 &  7.085E-1 & 1.445 &  7.991E-1 \\
 10.00 & 1.517 &  1.103E+0 & 1.340 &  1.381E+0 & 1.202 &  8.420E-1 & 1.518 &  9.175E-1 \\
 10.20 & 1.665 &  1.074E+0 & 1.540 &  1.383E+0 & 1.322 &  9.467E-1 & 1.624 &  1.020E+0 \\
 10.40 & 1.797 &  1.012E+0 & 1.711 &  1.351E+0 & 1.468 &  1.017E+0 & 1.772 &  1.097E+0 \\
 10.60 & 1.903 &  9.303E-1 & 1.874 &  1.293E+0 & 1.626 &  1.047E+0 & 1.949 &  1.128E+0 \\
 10.80 & 1.979 &  8.379E-1 & 1.976 &  1.209E+0 & 1.776 &  1.038E+0 & 2.123 &  1.104E+0 \\
 11.00 & 2.032 &  7.373E-1 & 2.043 &  1.113E+0 & 1.916 &  9.930E-1 & 2.282 &  1.031E+0 \\
 11.20 & 2.056 &  6.373E-1 & 2.089 &  9.976E-1 & 2.036 &  9.183E-1 & 2.409 &  9.190E-1 \\
 11.40 & 2.056 &  5.487E-1 & 2.105 &  8.619E-1 & 2.123 &  8.263E-1 & 2.488 &  7.913E-1 \\
 11.60 & 2.038 &  4.736E-1 & 2.108 &  7.305E-1 & 2.184 &  7.246E-1 & 2.528 &  6.625E-1 \\
 11.80 & 2.009 &  4.153E-1 & 2.100 &  6.147E-1 & 2.219 &  6.200E-1 & 2.537 &  5.420E-1 \\
 12.00 & 1.976 &  3.758E-1 & 2.080 &  5.093E-1 & 2.228 &  5.166E-1 & 2.520 &  4.364E-1 \\
 12.50 & 1.927 &  3.298E-1 & 1.979 &  3.180E-1 & 2.155 &  3.041E-1 & 2.426 &  2.545E-1 \\
 13.00 & 1.921 &  2.674E-1 & 1.860 &  2.677E-1 & 2.021 &  2.512E-1 & 2.321 &  1.598E-1 \\
 13.50 & 1.873 &  1.700E-1 & 1.797 &  2.679E-1 & 2.033 &  1.775E-1 & 2.228 &  1.079E-1 \\
 14.00 & 1.767 &  1.015E-1 & 1.749 &  2.330E-1 & 1.884 &  5.609E-2 & 2.144 &  7.781E-2 \\
 14.50 & 1.638 &  9.173E-2 & 1.647 &  2.033E-1 & 1.705 &  6.002E-2 & 2.059 &  6.304E-2 \\
 15.00 & 1.519 &  1.332E-1 & 1.520 &  2.355E-1 & 1.536 &  1.141E-1 & 1.970 &  6.498E-2 \\
 15.50 & 1.419 &  2.254E-1 & 1.414 &  3.297E-1 & 1.372 &  2.284E-1 & 1.876 &  8.932E-2 \\
 16.00 & 1.359 &  3.640E-1 & 1.369 &  5.008E-1 & 1.256 &  4.330E-1 & 1.782 &  1.435E-1 \\
 16.50 & 1.340 &  5.044E-1 & 1.397 &  7.162E-1 & 1.215 &  6.599E-1 & 1.699 &  2.325E-1 \\
 17.00 & 1.362 &  6.376E-1 & 1.455 &  8.875E-1 & 1.254 &  8.764E-1 & 1.643 &  3.569E-1 \\
 17.50 & 1.425 &  7.684E-1 & 1.534 &  1.030E+0 & 1.359 &  1.064E+0 & 1.633 &  5.032E-1 \\
 18.00 & 1.516 &  8.819E-1 & 1.615 &  1.131E+0 & 1.506 &  1.186E+0 & 1.674 &  6.459E-1 \\
 18.50 & 1.613 &  9.637E-1 & 1.694 &  1.201E+0 & 1.639 &  1.247E+0 & 1.753 &  7.651E-1 \\
 19.00 & 1.714 &  1.018E+0 & 1.774 &  1.252E+0 & 1.769 &  1.280E+0 & 1.859 &  8.567E-1 \\
 19.50 & 1.814 &  1.055E+0 & 1.845 &  1.271E+0 & 1.887 &  1.278E+0 & 1.957 &  9.221E-1 \\
 20.00 & 1.911 &  1.077E+0 & 1.913 &  1.276E+0 & 1.978 &  1.262E+0 & 2.053 &  9.610E-1 \\
 21.00 & 2.080 &  1.075E+0 & 2.051 &  1.250E+0 & 2.128 &  1.207E+0 & 2.251 &  9.523E-1 \\
 22.00 & 2.220 &  1.047E+0 & 2.148 &  1.198E+0 & 2.209 &  1.142E+0 & 2.389 &  8.883E-1 \\
 23.00 & 2.326 &  9.921E-1 & 2.214 &  1.156E+0 & 2.264 &  1.095E+0 & 2.468 &  7.984E-1 \\
 24.00 & 2.408 &  9.409E-1 & 2.264 &  1.120E+0 & 2.297 &  1.057E+0 & 2.488 &  7.291E-1 \\
 25.00 & 2.469 &  9.037E-1 & 2.303 &  1.095E+0 & 2.316 &  1.043E+0 & 2.484 &  6.716E-1 \\
 26.00 & 2.509 &  8.726E-1 & 2.339 &  1.080E+0 & 2.343 &  1.048E+0 & 2.462 &  6.343E-1 \\
 27.00 & 2.532 &  8.426E-1 & 2.378 &  1.067E+0 & 2.384 &  1.056E+0 & 2.441 &  6.249E-1 \\
 28.00 & 2.541 &  8.158E-1 & 2.416 &  1.053E+0 & 2.431 &  1.055E+0 & 2.431 &  6.325E-1 \\
 29.00 & 2.542 &  7.967E-1 & 2.450 &  1.040E+0 & 2.474 &  1.048E+0 & 2.431 &  6.471E-1 \\
 30.00 & 2.539 &  7.848E-1 & 2.483 &  1.029E+0 & 2.514 &  1.040E+0 & 2.442 &  6.653E-1 \\
\hline
\end{tabular}
}
\label{TabOptDat4}
\end{table*}

\addtocounter{table}{-1}
\begin{table*}
\caption{Continuation.} 

{\small
\begin{tabular}{rcccccccc}
\hline
\hline
\noalign{\smallskip}
$\lambda$ &  $n$  &    $k$    &  $n$  &    $k$    &  $n$  &    $k$    &  $n$  &    $k$    \\
\noalign{\smallskip}
\hline
\noalign{\smallskip}
 32.00 & 2.520 &  7.702E-1 & 2.550 &  1.007E+0 & 2.592 &  1.023E+0 & 2.492 &  6.986E-1 \\
 34.00 & 2.506 &  7.462E-1 & 2.618 &  9.793E-1 & 2.667 &  9.966E-1 & 2.559 &  7.022E-1 \\
 36.00 & 2.524 &  7.045E-1 & 2.677 &  9.416E-1 & 2.734 &  9.599E-1 & 2.617 &  6.910E-1 \\
 38.00 & 2.549 &  6.596E-1 & 2.722 &  9.001E-1 & 2.787 &  9.174E-1 & 2.671 &  6.748E-1 \\
 40.00 & 2.567 &  6.207E-1 & 2.757 &  8.606E-1 & 2.827 &  8.754E-1 & 2.723 &  6.488E-1 \\
 45.00 & 2.600 &  5.448E-1 & 2.815 &  7.827E-1 & 2.891 &  7.909E-1 & 2.808 &  5.601E-1 \\
 50.00 & 2.625 &  4.829E-1 & 2.856 &  7.268E-1 & 2.936 &  7.302E-1 & 2.847 &  4.853E-1 \\
 55.00 & 2.639 &  4.291E-1 & 2.890 &  6.814E-1 & 2.972 &  6.806E-1 & 2.872 &  4.260E-1 \\
 60.00 & 2.644 &  3.842E-1 & 2.917 &  6.423E-1 & 2.999 &  6.378E-1 & 2.889 &  3.736E-1 \\
 65.00 & 2.643 &  3.490E-1 & 2.935 &  6.090E-1 & 3.018 &  6.010E-1 & 2.896 &  3.274E-1 \\
 70.00 & 2.638 &  3.219E-1 & 2.948 &  5.820E-1 & 3.030 &  5.707E-1 & 2.896 &  2.890E-1 \\
 75.00 & 2.633 &  3.019E-1 & 2.956 &  5.610E-1 & 3.038 &  5.468E-1 & 2.891 &  2.581E-1 \\
 80.00 & 2.626 &  2.871E-1 & 2.962 &  5.453E-1 & 3.044 &  5.286E-1 & 2.884 &  2.342E-1 \\
100.00 & 2.606 &  2.631E-1 & 2.972 &  5.241E-1 & 3.050 &  4.995E-1 & 2.855 &  1.825E-1 \\
200.00 & 2.597 &  3.734E-1 & 3.036 &  7.532E-1 & 3.100 &  7.081E-1 & 2.811 &  1.944E-1 \\
\hline
\end{tabular}
}

\end{table*}

\end{appendix}


\listofobjects

\end{document}